\documentclass[sigconf,nonacm]{acmart}

\usepackage{pvldb}
\renewcommand\vldbdoi{XX.XX/XXX.XX}
\renewcommand\vldbpages{XXX-XXX}

\renewcommand\vldbavailabilityurl{https://github.com/Sohaib03/carousel_artifact}

\renewcommand\vldbpagestyle{plain}

\usepackage{xcolor} 
\usepackage{balance}

\usepackage{subcaption}
\usepackage{amsthm}

\usepackage{graphicx}   
\usepackage{subcaption} 
\usepackage{listings}
\lstdefinestyle{promptstyle}{
    language={}, 
    basicstyle=\ttfamily\small, 
    breaklines=true, 
    columns=fullflexible, 
    frame=single, 
    frameround=tttt, 
    backgroundcolor=\color[rgb]{0.95,0.95,0.95}, 
    numbersep=5pt, 
    showstringspaces=false, 
    rulecolor=\color{black},
}

\usepackage{caption}
\DeclareCaptionType{promptblock}[Prompt Block]

\DeclareCaptionType{queryblock}[Query Block]

\usepackage{booktabs}
\usepackage{multirow}

\usepackage{tikz}
\usepackage{amsmath}
\usepackage{xcolor}
\usepackage{algpseudocode}

\usetikzlibrary{matrix, positioning, decorations.pathreplacing, calc, arrows.meta, fit}

\usepackage{algorithm}      
\usepackage{algpseudocode}  
\usepackage{amsmath}        

\usepackage{pgfplots}
\pgfplotsset{compat=1.17}

\usepackage{svg}
\usepackage{graphicx}
\usepackage{algorithm}
\usepackage{algpseudocode}
\usepackage{tikz}
\usepackage{pifont}
\usetikzlibrary{positioning,calc,arrows.meta,shapes.geometric,shapes.arrows,fit,backgrounds,
                decorations.pathreplacing,patterns,shapes.misc}

\definecolor{cBlue}{HTML}{DAE8F5}      \definecolor{cBlueD}{HTML}{2E5C8A}
\definecolor{cOrange}{HTML}{FBE5D5}    \definecolor{cOrangeD}{HTML}{C55A11}
\definecolor{cPink}{HTML}{F5DEDE}      \definecolor{cPinkD}{HTML}{A64D4D}
\definecolor{cPurple}{HTML}{E5DFEC}    \definecolor{cPurpleD}{HTML}{7030A0}
\definecolor{cGreen}{HTML}{E2EFDA}     \definecolor{cGreenD}{HTML}{548235}
\definecolor{cHi}{HTML}{BDD7EE}        \definecolor{cGrey}{HTML}{F2F2F2}
\definecolor{cGreyD}{HTML}{7F7F7F}     \definecolor{cRed}{HTML}{C00000}
\definecolor{cYellow}{HTML}{FFF2CC}    \definecolor{cYellowD}{HTML}{BF8F00}

\tikzset{
  panel/.style        = {draw=black, line width=0.8pt, rounded corners=1.5pt,
                         inner sep=5pt},
  grouped/.style      = {draw=none, fill=cGrey, rounded corners=7pt,
                         inner sep=6pt},
  ptitle/.style       = {font=\bfseries\footnotesize, align=center},
  plabel/.style       = {font=\scriptsize, align=center},
  tiny/.style         = {font=\tiny, align=center},
  ttiny/.style        = {font=\ttfamily\tiny, align=center},
  cell/.style         = {draw=black, line width=0.5pt, minimum width=6mm,
                         minimum height=4mm, inner sep=1pt, font=\ttfamily\tiny},
  bluecell/.style     = {cell, fill=cBlue},
  orangecell/.style   = {cell, fill=cOrange},
  purplecell/.style   = {cell, fill=cPurple},
  greencell/.style    = {cell, fill=cGreen},
  hicell/.style       = {cell, fill=cHi},
  dpoint/.style       = {circle, draw=cGreyD, fill=cGrey, minimum size=1.6mm,
                         inner sep=0pt, line width=0.4pt},
  qpoint/.style       = {circle, draw=cGreenD, fill=cGreenD, minimum size=2.2mm,
                         inner sep=0pt},
  rpoint/.style       = {circle, draw=cRed, fill=cRed, minimum size=2.2mm,
                         inner sep=0pt},
  flow/.style         = {-{Stealth[length=2.2mm,width=1.8mm]}, line width=0.7pt},
  thinflow/.style     = {-{Stealth[length=1.6mm,width=1.3mm]}, line width=0.4pt,
                         draw=cGreyD},
  blockarrow/.style   = {single arrow, draw=black, fill=cBlue, line width=0.6pt,
                         minimum height=6mm, minimum width=4mm,
                         single arrow head extend=1mm, inner sep=1.5pt},
  stagenum/.style     = {circle, draw=black, fill=white, line width=0.6pt,
                         inner sep=0.5pt, font=\bfseries\tiny, minimum size=3.4mm},
}

\definecolor{customDarkOrange}{HTML}{FD9E54}
\definecolor{customLightOrange}{HTML}{FDD3B4}
\definecolor{customLightGray}{HTML}{E0E0E0}
\definecolor{customLightBlue}{HTML}{A3CDF0}

\definecolor{d0f}{HTML}{CECBF6}\definecolor{d0e}{HTML}{534AB7}\definecolor{d0t}{HTML}{26215C}
\definecolor{d1f}{HTML}{FAC775}\definecolor{d1e}{HTML}{854F0B}\definecolor{d1t}{HTML}{412402}
\definecolor{nf}{HTML}{EAF1FB}\definecolor{ne}{HTML}{B5D4F4}\definecolor{nt}{HTML}{2E4A7F}
\definecolor{qf}{HTML}{EAF3DE}\definecolor{qe}{HTML}{639922}\definecolor{qt}{HTML}{27500A}
\definecolor{accf}{HTML}{FAECE7}\definecolor{ctedge}{HTML}{993C1D}\definecolor{acct}{HTML}{712B13}
\definecolor{lbl}{HTML}{5F5E5A}

\usepackage{eso-pic}

\newcommand{\distributionstatement}{
    Approved for Public Release on 07 Oct 2026. Distribution Is Unlimited; Case Number: AFRL-2026-4302.
}

\AddToShipoutPictureFG{%
  \AtPageLowerLeft{%
    \raisebox{5mm}[0pt][0pt]{%
      \makebox[\paperwidth][c]{%
        \scriptsize\distributionstatement
      }%
    }%
  }%
}

\begin{document}

\title{CAROUSEL: GPU-Accelerated Private Vector Search via Homomorphic Sketching}

\author{Sohaib}
\email{sohaib@ucsb.edu}
\affiliation{%
  \institution{UC Santa Barbara}
  \country{}
}

\author{Divyakant Agrawal}
\email{divyagrawal@ucsb.edu}
\affiliation{%
  \institution{UC Santa Barbara}
  \country{}
}

\author{Amr El Abbadi}
\email{amr@cs.ucsb.edu}
\affiliation{%
  \institution{UC Santa Barbara}
  \country{}
}

\author{Soamar Homsi}
\email{soamar.homsi@us.af.mil}
\affiliation{%
  \institution{Air Force Research Laboratory}
  \country{}
}

\begin{abstract}

Vector search on untrusted infrastructure exposes the \emph{query embedding}, a faithful summary of the user's intent. Fully homomorphic encryption can hide the query, but existing approaches still require a full scan followed by ranking of encrypted scores. Scalable private systems avoid encrypted ranking by either sending scores from a single cluster to the client or moving the search to the client, which must first stream the entire vector index for preprocessing. We present \textsc{Carousel}, which instead scores the full corpus under encryption and compresses the results for client-side ranking. The server scores every vector under encryption, raises the scores to a power that preserves their order by magnitude while suppressing all but a handful, and compresses the resulting sparse vector into a logarithmic number of ciphertexts using homomorphic additions alone. Because every vector is scored, \textsc{Carousel} incurs no recall loss from discarding candidates. Since clients hold no database-dependent state, insertions and deletions take effect immediately at the cost of a single vector write. \textsc{Carousel} searches $1$ billion vectors of BIGANN SIFT1B across multiple GPUs and achieves recall@100 of $0.9715$. At $100$ million SIFT vectors, it achieves recall@100 of $0.9845$ with a query latency of $3.735$ seconds on a single GPU.

\end{abstract}

\maketitle

\pagestyle{plain}

\section{Introduction}
\label{sec:intro}

Modern information retrieval systems increasingly rely on vector similarity search as their principal operation. Documents, images, code, and product catalogs are embedded into a high-dimensional space in which
semantic similarity is inner-product proximity, and a query is answered by finding the stored vectors nearest to the query's embedding. Search engines, recommendation systems, and retrieval-augmented generation pipelines all depend on this primitive, and vector databases serving hundreds of millions of embeddings are now routine infrastructure~\cite{milvus}. The first step of every such query, however, is to hand the query embedding to the server and that embedding is a faithful summary of what the user wants to know. This is a privacy risk as search queries reveal sensitive personal information such as where they are, what they are doing, or what they are living through~\cite{aol,hafner}. 

Efficient vector search algorithms~\cite{malkov2020efficient} typically rely on query dependent access patterns, which requires that the query remain in plaintext. For a search to be completely private, the server's actions must be completely independent of the query itself. This means the server cannot skip over any part of its index, because the decision to skip would reveal information about the search i.e. which part of the index were not relevant to the query. Private search therefore forces a linear scan on every
request, and recent literature is organized around either paying for that scan
or avoiding it by preprocessing. Every approach considered so far in the literature costs something that the user notices.

These costs usually appear as either massive communication overhead or reduced search accuracy. Systems that offer the strongest privacy, where the server learns nothing at all about the query, face this tradeoff clearly. For example, Tiptoe ~\cite{tiptoe} suffers from a drop in search quality, while PACMANN~\cite{pacmann} relies on heavy preprocessing that requires the server to stream the whole index to the client. Even relaxing the privacy guarantees does not solve the problem. Wally~\cite{wally}, which uses weaker differential privacy, does not improve search quality significantly. Thus, the fundamental challenge remains unsolved, providing complete privacy and high search accuracy without a heavy communication cost.

To resolve this challenge, this paper presents \textsc{Carousel}, a GPU-accelerated vector search system that achieves complete privacy with high search accuracy and low communication overhead. The tradeoff prior systems face is not strictly necessary. Rather, it arises from how they choose to reduce communication overhead. The fundamental problem is that an exhaustive scan produces a relevance score for every document, creating a response far too large for transmission. Prior systems (Tiptoe, Wally) address this by pruning the database before the query arrives, estimating which documents are relevant and this filtering causes loss of accuracy. \textsc{Carousel} avoids this by scoring every document and compressing only the final answer, discarding nothing before ranking is complete.

\textsc{Carousel} executes this secure scan using Fully Homomorphic Encryption (FHE)~\cite{gentry, ckks}, which enables the server to perform additions and multiplications directly on ciphertexts. Selecting the best matches from the resulting scores, however, requires comparison, which FHE does not support natively. A single comparison must be approximated by a high-degree polynomial, making the sort needed to rank even a few hundred encrypted scores cost seconds~\cite{mazzone}. Avoiding this sort leaves the server unable to select the top matches, so it would have to return every score for the client to rank. To overcome the bottleneck of returning millions of encrypted relevance scores, \textsc{Carousel} employs a \emph{sketch}, a compact summary of a large number of vectors that preserves its largest entries for later recovery. However, sketching techniques fundamentally rely on data sparsity, assuming that the vast majority of entries are near zero. A list of relevance scores violates this assumption as most documents exhibit a modest, non-zero similarity to the query, and even the top matches sit only marginally higher than this baseline. 

To make the score list sparse without altering which documents rank highest, we introduce a step called \emph{sharpening}. The server raises every encrypted score to a power, preserving their ordering by absolute magnitude. Since all scores fall between -1 and 1, larger values shrink far more slowly than smaller ones. After this operation, the broad middle of poor-to-modest matches collapses rapidly toward zero, while the highest scores retain their magnitude. This widens the numerical gap between the top matches and the rest, allowing a sketch to compress the results while preserving high retrieval accuracy.

The sketch itself is constructed using only addition, the cheapest operation FHE offers. The server compacts the sparse scores into a few compact summary rows and returns those in place of the millions of initial scores. The client then decrypts the rows and recovers the top matches locally. Therefore, every comparison in the system occurs on the client after decryption. This ensures that the server never has to compare or sort encrypted values, avoiding the step that is expensive under encryption.

The matches the client recovers, however, are positions in the index rather than the documents themselves. To fetch the indexed data without revealing which the client wants, it uses \emph{private information retrieval} (PIR), a protocol that allows a client to read an entry from a server-held database while the server learns nothing about which entry was read~\cite{simplepir, piano, hintlesspir}. \textsc{Carousel} addresses only the ranking step and composes with any PIR scheme for the retrieval that follows. The two together form an end-to-end private pipeline, the query stays hidden during search, and the retrieved documents stay hidden during retrieval.

\textsc{Carousel}’s design also enables efficient online updates. It maintains a flat array of vectors, without a clustering or graph index to update. Insertions append a vector, and deletions zero its coordinates. When encoded diagonals are cached, only the affected blocks need to be refreshed. These changes require no client-side refresh of corpus-dependent search metadata. Tiptoe ~\cite{tiptoe} describes support for continuous updates by assigning new or modified documents to clusters and distributing updated centroids and metadata to clients, but does not evaluate update latency or query performance under concurrent updates. We evaluate both the cost of insertions and deletions and their impact on query latency under sustained update workloads.

We implement \textsc{Carousel} using CKKS~\cite{ckks}, a fully homomorphic encryption scheme, and support both single- and multi-GPU execution. On MS MARCO, \textsc{Carousel} reaches an MRR@100 of $0.303$, matching a non-private graph index and exceeding Tiptoe at $0.13$, Wally at $0.18$, and Pacmann at $0.266$. On SIFT-100M, it achieves recall@10 of $0.9785$, compared with $0.90$ for Pacmann, and completes each query in $3.735$ seconds on a single GPU. On BIGANN SIFT1B, \textsc{Carousel} searches all $1$ billion vectors across multiple GPUs and achieves recall@100 of $0.9715$.

\textsc{Carousel} makes the following contributions:
\begin{itemize}
\item A private vector-search design that compresses the answer with a sketch rather than pruning the database, matching non-private search quality with logarithmic communication and cryptographic query privacy.
\item An encrypted pipeline that sparsifies scores through power-based sharpening and moves all ranking to the client, avoiding encrypted comparison.
\item A dynamic, GPU-accelerated implementation that supports single-write updates and multi-GPU execution, evaluated on SIFT, MS MARCO, and GIST at scales up to $10^9$ vectors.
\end{itemize}

\section{Background and Preliminaries}
\label{sec:background}

\subsection{Private Vector Search}
\label{sec:background:problem}

A vector search index holds a database $\mathcal{D} = \{v_1, \dots, v_N\}$ of $N$
vectors in $\mathbb{R}^{d}$, obtained by passing documents, images, or other content
through a pretrained embedding model. A client holding a query vector
$q \in \mathbb{R}^{d}$ wants the identifiers of the $k$ database vectors most similar
to $q$, which for the embedding models used in this paper means largest inner product
$\langle q, v_i\rangle$. All vectors are normalized to unit length, so the inner
product coincides with cosine similarity and lies in $[-1,1]$.

In the private version of this problem, the server must learn absolutely nothing about the query, $q$. We guarantee this using standard semantic security for the query. Our threat model assumes a single {\emph{semi-honest} server that follows the protocol, but still attempts to deduce the query from its observations.

\begin{definition}[Query privacy]
\label{def:privacy}
Let $\mathsf{View}_{\mathcal{S}^*}[\mathcal{C}(q)]$ denote the view of a server
$\mathcal{S}^*$ in its interaction with a client that searches for $q$. A private
vector search protocol satisfies query privacy if, for every
$\mathcal{S}^*$ and every pair of queries $q_0, q_1$, the distributions
$\mathsf{View}_{\mathcal{S}^*}[\mathcal{C}(q_0)]$ and
$\mathsf{View}_{\mathcal{S}^*}[\mathcal{C}(q_1)]$ are computationally
indistinguishable.
\end{definition}

The view includes every message the server receives, the memory it touches, and
the sizes and timing of the responses it returns. Definition~\ref{def:privacy}
forces all of these to be independent of the query, which is the formal statement
of the linear scan requirement discussed in Section~\ref{sec:intro}.
Table~\ref{tab:notation} collects the notation used throughout.

\begin{table}[t]
\centering
\small
\caption{Notation.}
\label{tab:notation}
\begin{tabular}{ll@{\qquad}ll}
\toprule
Sym. & Meaning & Sym. & Meaning \\
\midrule
$N$   & database vectors            & $n$      & CKKS ring dimension \\
$d$   & vector dimension            & $\ell$   & slots, $\ell = n/2$ \\
$k$   & results returned            & $L$      & levels in the chain \\
$q, v_i$ & query and $i$th vector   & $\Delta$ & CKKS scaling factor \\
$s_i$ & score of $v_i$ against $q$  & $p$      & sharpening exponent \\
$B$   & blocks, $\lceil N/\ell \rceil$ & $M$   & fold matrix \\
$R$   & fold rows returned             &   &  \\
\bottomrule
\end{tabular}
\end{table}

\subsection{Fully Homomorphic Encryption}
\label{sec:background:fhe}

Fully homomorphic encryption (FHE) allows a server to compute on encrypted data without decrypting it. Given an encryption of $m$ and public evaluation keys, the server can evaluate a supported function $f$ to obtain an encryption of $f(m)$, without interacting with the key holder. Many practical schemes derive their security from the ring learning with errors assumption~\cite{lpr}.

We use CKKS~\cite{ckks}, which supports approximate arithmetic on real or complex numbers and is well suited to computing similarity scores. A fresh ciphertext consists of two polynomials in the ring $R_Q = \mathbb{Z}_Q[X]/(X^n+1)$, where $n$ is a power of two and $Q$ is the ciphertext modulus. CKKS packs up to $\ell = n/2$ values into the \emph{slots} of a single plaintext polynomial, allowing additions and multiplications to operate on all slots simultaneously in a SIMD manner. During encoding, a scaling factor $\Delta$ controls the precision with which these values are represented in the integer polynomial ring. Decryption and decoding recover approximate values, and homomorphic operations introduce and propagate numerical errors. The encryption parameters must therefore provide sufficient precision for the intended computation.

CKKS supports a bounded number of multiplication stages through a \emph{modulus chain}, $Q = q_0 q_1 \cdots q_L$. Multiplying two ciphertexts at scale $\Delta$ produces a ciphertext at scale $\Delta^2$. A subsequent \emph{rescaling} operation divides the scale and modulus by one prime, bringing the scale back near $\Delta$ and consuming one level. The circuit's \emph{multiplicative depth}, the largest number of sequential multiplication stages along any computation path, must therefore fit within the available level budget. Longer computations require \emph{bootstrapping} to refresh the ciphertext and restore usable levels, at substantial computational cost. We design \textsc{Carousel} to complete its encrypted computation within the initial level budget, avoiding bootstrapping entirely.

Table~\ref{tab:ops} lists all the standard homomorphic operators we use in our system, along with their required keys and relative computation costs. These costs vary widely and this variation directly shapes how we design our pipeline. For example, rotation is highly expensive because it relies on key switching, which dominates both runtime and memory. Every distinct rotation shift requires the server to hold a separate, multi-megabyte evaluation key. In our setup, each key is about 10 MB, so minimizing the total number of rotation keys is a critical goal. Finally, computation cost scales with the size of the ciphertext, meaning it is always cheaper to compute at a lower multiplicative level than at a higher one.

\begin{table}[h]
\centering\small
\caption{Homomorphic operators, the key material each requires, the levels it
consumes, and its relative cost.}
\label{tab:ops}
\begin{tabular}{llcl}
\toprule
operator & key & levels & cost \\
\midrule
\texttt{Add}$(\mathsf{ct}, \mathsf{ct}/\mathsf{pt})$ & --- & $0$ & negligible \\
\texttt{MulPt}$(\mathsf{ct}, \mathsf{pt})$ & --- & $0$ & low \\
\texttt{Mul}$(\mathsf{ct}_0, \mathsf{ct}_1)$ & --- & $0$ & moderate \\
\texttt{Relin}$(\mathsf{ct}, \mathbf{RK})$ & relin. & $0$ & high \\
\texttt{Rescale}$(\mathsf{ct})$ & --- & $1$ & low \\
\texttt{Rot}$(\mathsf{ct}, \mathbf{RK}_{\mathit{step}})$ & one per shift & $0$ & high \\
\texttt{Conj}$(\mathsf{ct}, \mathbf{CK})$ & conj. & $0$ & high \\
\bottomrule
\end{tabular}
\end{table}

The asymmetry in computational costs among FHE operators introduces two fundamental constraints that shape our system architecture. First, an operator
built exclusively from \texttt{Add} and \texttt{MulPt} consumes no evaluation
keys, adds no depth, and introduces no key switching error, which makes it
qualitatively cheaper than one that uses rotations. Second, comparison is not a native FHE operation.
Sign, minimum, maximum and argmax must be approximated by compositions of high
degree polynomials~\cite{cheoncomp}, each consuming several levels, which is why
selecting the top $k$ entries of an encrypted vector is expensive even for a few
hundred candidates~\cite{mazzone}. This is the reason \textsc{Carousel}
performs no comparison under encryption.

\subsection{Prior Approaches to Private Vector Search}
\label{sec:background:related}

Private nearest-neighbor search splits into two families. Interactive protocols built on secure multi-party computation can exploit an index and so avoid touching the whole database, at the price of many rounds and of trust assumptions such as non-colluding servers. Non-interactive protocols built on homomorphic encryption need one round and one server, but must compute over every record. \textsc{Carousel} belongs to the second family.

\smallskip\noindent\textbf{Private search over a public corpus.}
Our design directly compares to existing architectures that hide only the query while
relying on a public corpus and a single untrusted server. Coeus~\cite{coeus} first
posed the problem for text, privately scoring Wikipedia articles with tf-idf at a cost
that does not reach web scale. Tiptoe~\cite{tiptoe} replaced tf-idf with semantic
embeddings, reducing private search to private nearest-neighbor search, and answered it
with a preprocessing-friendly linearly homomorphic scheme~\cite{simplepir} together
with clustering to bound the response size. Clustering partitions the corpus offline,
typically by $k$-means, which groups vectors around a set of centroids so that each
vector belongs to the centroid nearest it. At query time the client probes only the cluster whose centroid is closest to its query, so the server effectively scores a fraction of the corpus rather than all $N$. Tiptoe~\cite{tiptoe} uses $\sqrt{N}$ clusters, and a relevant document outside the probed cluster cannot be retrieved. Wally~\cite{wally} keeps the clustering and relaxes the guarantee instead, revealing the probed cluster under differential privacy using cover queries routed through an anonymity network. Its overhead falls as the concurrent client population grows but it requires an anonymity network to strip identifiers and shuffle requests. Pacmann~\cite{pacmann} moves graph traversal to the client. Starting from an entry node, the client repeatedly uses preprocessing PIR~\cite{piano} to fetch the neighbor list of the node closest to the query until the search converges. This shifts most computation and state to the client and before querying, it must stream the entire graph index and store $0.6$~GB of hints for three million vectors or $2.9$~GB for one hundred million. Each query also requires $20$--$30$ sequential network rounds. These three systems form our baselines.

\smallskip\noindent\textbf{Other private nearest-neighbor protocols.}
Another line of work uses interactive multi-party computation (MPC).
SANNS~\cite{sanns} combines distributed oblivious RAM and garbled circuits, while
Panther~\cite{panther} reduces communication through batch PIR. Both use a single server and prevent the client from learning anything about the corpus beyond the search results. Servan-Schreiber et al.~\cite{preco}
instead require two non-colluding servers to obtain sublinear communication.
Compass~\cite{compass} considers a different setting, where the client encrypts
and uploads its own corpus, protecting it from the server.
GraSS~\cite{grass}, the first fully homomorphic graph index, still takes hours per query on datasets much smaller than ours. Because of this cost, prior work viewed fully homomorphic scans as impractical. HANNS~\cite{hanns} also assumes that non-interactive search requires a full scan, but lowers its cost by ordering clusters to reduce encrypted comparisons and quantizing vectors to reduce storage.





\section{The \textsc{Carousel} Query Pipeline}
\label{sec:design}

\subsection{Overview}
\label{sec:design:overview}

The server holds a database of $N$ vectors $v_1, \dots, v_N$ in $\mathbb{R}^{d}$, and a client holds a query $q \in \mathbb{R}^{d}$. The client wants to retrieve the indices of the $k$ database vectors most similar to $q$, where similarity is the inner product $\langle q, v_i \rangle$, and must learn them without revealing $q$ to the server. Once the indices are retrieved, the client can use PIR techniques to retrieve the corresponding records. All vectors are normalised to unit length, so this inner product is the cosine similarity and lies in $[-1,1]$. \textsc{Carousel} solves this with fully homomorphic encryption where the server computes on the encrypted query and the scores stay encrypted as well.

\textsc{Carousel} answers a query in four stages. The first three run on the server under encryption but the last runs on the client after decryption.

\begin{itemize}
  \item \textbf{Scan.} The server computes the inner product of the encrypted query
  against every database vector, producing an encrypted score $s_i = \langle q, v_i
  \rangle$ for every $i$.
  \item \textbf{Sharpen.} The server raises each encrypted score to a power,
  $s_i \mapsto s_i^{p}$. This preserves the ranking of the strong matches while driving
  the rest toward zero, turning a dense score vector into a sparse one.
  \item \textbf{Fold.} The server compresses the sparse scores into $R$ summary rows
  using a fixed binary matrix $M$, needing only homomorphic addition:
  $\mathrm{row}_r = \sum_{i} M_{r,i}\, s_i^{p}$.
  \item \textbf{Decode.} The client decrypts the $R$ rows and recovers the top-$k$
  identifiers locally. Every comparison in the system happens here, in plaintext.
\end{itemize}

The scores of $\ell$ consecutive vectors are packed into the slots of a single ciphertext. We call such a group of $\ell$ vectors a \emph{block}, so a corpus of $N$ vectors occupies $B = \lceil N/\ell \rceil$ blocks.
We illustrate the pipeline throughout with a corpus of $N = 64$ vectors of dimension $d = 4$ and a ciphertext of $\ell = 8$ slots, so the similarity scores for a given query occupies $B = 8$ blocks.


\subsection{Data Layout}
\label{sec:design:layout}

The client encrypts its query into a single ciphertext. The dimension $d$ of a query (and the vectors in the database) is much smaller than the number of slots $\ell$ in a ciphertext. Instead of placing the query once and leaving the remaining slots empty, the client tiles the query across all
slots, repeating its $d$ coordinates $\ell/d$ times. Figure~\ref{fig:tiling} shows a query with $d = 4$ coordinates and a ciphertext has $\ell = 8$ slots, so the pattern $q_0, q_1, q_2, q_3$ appears twice where $q_i$ refers to the $i$-th
coordinate of the query vector $q$.
%
%
\begin{figure}[h]
\centering
\resizebox{\columnwidth}{!}{%
\begin{tikzpicture}[x=0.62cm, y=0.74cm,
  /utils/exec={%
    \definecolor{qf}{HTML}{EAF3DE}\definecolor{qe}{HTML}{639922}%
    \definecolor{qt}{HTML}{27500A}\definecolor{ctedge}{HTML}{993C1D}%
    \definecolor{d1e}{HTML}{854F0B}\definecolor{d1t}{HTML}{412402}%
    \definecolor{lbl}{HTML}{5F5E5A}},
  q/.style   ={minimum width=0.56cm, minimum height=0.62cm, inner sep=0pt,
               line width=0.7pt, font=\scriptsize, draw=qe, fill=qf, text=qt},
  lab/.style ={text=lbl, font=\small},
  dlab/.style={font=\small\bfseries, text=d1t},
  foot/.style={text=lbl, font=\footnotesize},
]

\def\sx{1.55}          
\def\qdim{4}           
\def\lft{1.00}         
\def\rgt{9.10}         

\newcommand{\period}[4]{%
  \draw[draw=lbl, line width=0.6pt]
    (#1,#3) -- (#1,#3-0.16) -- (#2,#3-0.16) -- (#2,#3);
  \node[dlab, anchor=north, font=\scriptsize] at ({(#1+#2)/2},#3-0.30) {#4};}

\foreach \s in {0,...,7}{\node[lab, font=\scriptsize] at (\sx+\s,0.95) {\s};}
\node[lab, anchor=east, font=\scriptsize] at (\sx-0.62,0.95) {slot};

\node[lab, anchor=east] at (\sx-0.62,0) {$\mathsf{ct}(q)$};
\foreach \s in {0,...,7}{%
  \pgfmathtruncatemacro{\v}{mod(\s,\qdim)}%
  \node[q] at (\sx+\s,0) {$q_{\v}$};}
\draw[draw=ctedge, line width=1.1pt, rounded corners=3pt]
  (\lft,-0.52) rectangle (\rgt,0.52);
\draw[draw=d1e, line width=0.9pt, dash pattern=on 1.8pt off 1.6pt]
  (\sx+3.5,0.40) -- (\sx+3.5,-0.40);
\node[lab, anchor=west] at (\rgt+0.55,0) {$d'=4$, \ $\ell=8$};

\period{\lft}{\sx+3.45}{-0.62}{copy $0$}
\period{\sx+3.55}{\rgt}{-0.62}{copy $1$}

\node[foot] at (\sx+3.5,-1.80)
  {slot $j$ holds $q_{j \bmod d'}$: \ $\ell/d'=2$ copies, and no slot sits idle};

\end{tikzpicture}}
\caption{A four-dimensional query packed into a ciphertext with eight slots,
tiled twice to fill all slots.}
\label{fig:tiling}
\end{figure}
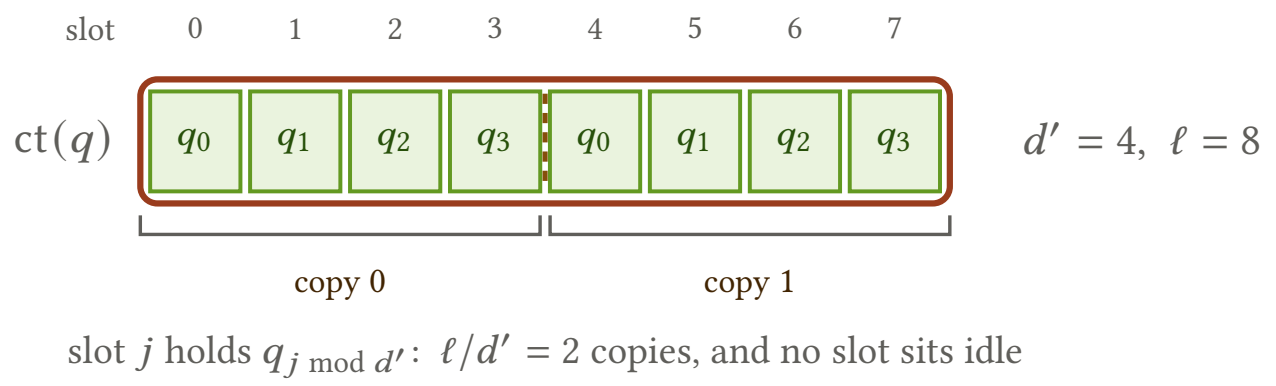

\begin{figure}[t]
\centering
\resizebox{\columnwidth}{!}{%
\begin{tikzpicture}[x=0.62cm, y=0.74cm,
  /utils/exec={%
    \definecolor{d0f}{HTML}{CECBF6}\definecolor{d0e}{HTML}{534AB7}%
    \definecolor{d0t}{HTML}{26215C}\definecolor{d1f}{HTML}{FAC775}%
    \definecolor{d1e}{HTML}{854F0B}\definecolor{d1t}{HTML}{412402}%
    \definecolor{qf}{HTML}{EAF3DE}\definecolor{qe}{HTML}{639922}%
    \definecolor{qt}{HTML}{27500A}\definecolor{ctedge}{HTML}{993C1D}%
    \definecolor{lbl}{HTML}{5F5E5A}},
  cell/.style ={minimum width=0.87cm, minimum height=0.64cm, inner sep=0pt,
                line width=0.7pt, font=\scriptsize},
  q/.style    ={cell, draw=qe, fill=qf, text=qt},
  lab/.style  ={text=lbl, font=\small},
  hdr/.style  ={text=lbl, font=\small\itshape},
  foot/.style ={text=lbl, font=\footnotesize},
]

\def\pit{1.5}                        
\def\bL{-0.82}\def\bR{11.32}         
\def\bC{5.25}                        
\def\hy{0.53}                        

\def\fillof#1{\ifcase#1 d0f\or d1f\fi}
\def\edgeof#1{\ifcase#1 d0e\or d1e\fi}
\def\textof#1{\ifcase#1 d0t\or d1t\fi}

\newcommand{\group}[4]{%
  \draw[draw=lbl, line width=0.6pt]
    (#1,#3) -- (#1,#3-0.16) -- (#2,#3-0.16) -- (#2,#3);
  \node[lab, anchor=north, font=\scriptsize] at ({(#1+#2)/2},#3-0.28) {#4};}

\foreach \s in {0,...,7}{\node[lab, font=\scriptsize] at (\s*\pit,1.15) {\s};}
\node[lab, font=\scriptsize, anchor=east] at (\bL-0.25,1.15) {slot};

\foreach \s in {0,...,7}{%
  \pgfmathtruncatemacro{\m}{mod(\s,4)}%
  \node[q] at (\s*\pit,0) {$q_{\m}$};}
\draw[draw=ctedge, line width=1.1pt, rounded corners=3pt]
  (\bL,-\hy) rectangle (\bR,\hy);
\node[lab, anchor=east] at (\bL-0.25,0) {$\mathsf{ct}(q)$};

\node[lab, font=\normalsize, anchor=east] at (\bL-0.25,-0.75) {$\odot$};

\foreach \s in {0,...,7}{%
  \pgfmathtruncatemacro{\vv}{int(\s/4)}%
  \pgfmathtruncatemacro{\m}{mod(\s,4)}%
  \node[cell, fill=\fillof{\vv}, draw=\edgeof{\vv}, text=\textof{\vv}]
    at (\s*\pit,-1.5) {$v_{\vv,\m}$};}
\draw[draw=lbl, dash pattern=on 2pt off 1.6pt, line width=0.7pt, rounded corners=3pt]
  (\bL,-1.5-\hy) rectangle (\bR,-1.5+\hy);
\node[lab, anchor=east] at (\bL-0.25,-1.5) {$\mathsf{pt}(V)$};

\draw[line width=1pt, draw=lbl] (\bL,-2.45) -- (\bR,-2.45);
\foreach \s in {0,...,7}{%
  \pgfmathtruncatemacro{\vv}{int(\s/4)}%
  \pgfmathtruncatemacro{\m}{mod(\s,4)}%
  \node[cell, fill=\fillof{\vv}, draw=\edgeof{\vv}, text=\textof{\vv}]
    at (\s*\pit,-3.35) {$q_{\m}v_{\vv,\m}$};}
\draw[draw=ctedge, line width=1.1pt, rounded corners=3pt]
  (\bL,-3.35-\hy) rectangle (\bR,-3.35+\hy);
\node[lab, anchor=east] at (\bL-0.25,-3.35) {product};

\group{\bL}{4.67}{-3.95}{sum these four $\to\ \langle q,v_0\rangle$}
\group{5.93}{\bR}{-3.95}{sum these four $\to\ \langle q,v_1\rangle$}


\end{tikzpicture}}
\caption{Row-major layout. The tiled query multiplies elementwise against two
vectors laid out in consecutive slots.
}
\label{fig:rowmajor}
\end{figure}
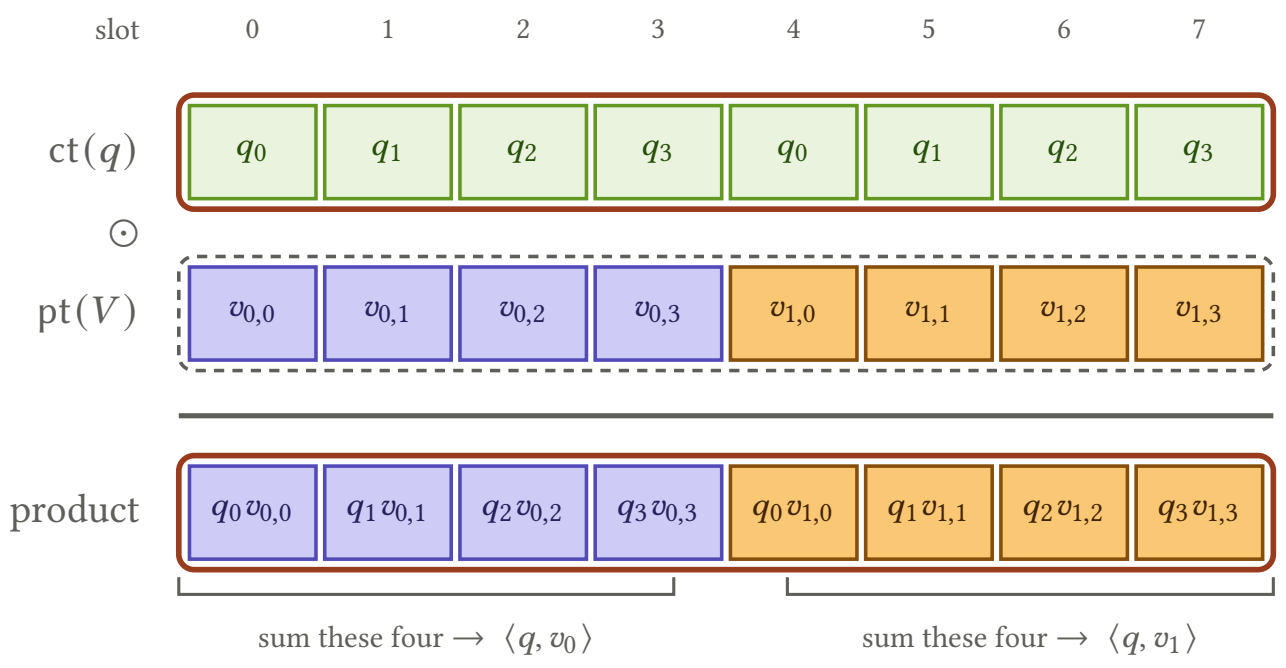

The corpus is laid out to match this tiled query. CKKS multiplies two operands slot by slot, so a product is formed only between values in corresponding slots. Consider storing the corpus in row-major order, with each vector's $d$ coordinates placed contiguously across slots. Figure~\ref{fig:rowmajor} shows what this produces where $v_{i,j}$ refers to the $j$-th coordinate of the $i$-th vector, $v_i$. 
Slot $0$ holds $q_0$ against $v_{0,0}$, slot $1$ holds $q_1$ against $v_{0,1}$, and slot $4$ holds $q_0$ against $v_{1,0}$. Every slot carries a product from a different vector coordinate, and the $d$ products belonging to one vector are scattered across $d$ separate slots. Recovering a single inner product would require summing these 4 slots together, and summing within a ciphertext requires rotations, which need rotation keys and are computationally expensive. 

To avoid this limitation, we adapt the Halevi--Shoup diagonal representation for homomorphic matrix--vector multiplication~\cite{halevi2014algorithms}. Instead of storing vectors in row-major order, we build $d$ plaintext diagonals, each containing one coordinate from every vector in the block. Diagonal $k$ is
\begin{equation}
  \mathrm{diag}[k][i] \;=\; v_i\big[(i{+}k)\bmod d\big],
  \label{eq:diag}
\end{equation}
so slot $i$ carries one coordinate of vector $i$. The offset by $i$ staggers which coordinate each slot takes, which makes a one-slot shift of the query line up with the next diagonal, thus simplifying the product and summation operations and avoiding expensive and unnecessary rotations. Figure~\ref{fig:diag} shows all four diagonals of a block of eight four-dimensional vectors, each filling all $\ell = 8$ slots.

\begin{algorithm}[t]
\caption{Server processing of one block $b$}
\label{alg:server}
\begin{algorithmic}[1]
\Require rotated query $\mathsf{RQ}[0..d{-}1]$, fold matrix $M$,
         row accumulators $\mathsf{rows}[0..R{-}1]$
\For{$k = 0$ to $d-1$}
  \State $\mathsf{DIAG}[k] \gets \textsc{Encode}(v_k)$
         \Comment{Eq.~\ref{eq:diag}}
\EndFor
\State $\mathsf{acc} \gets \mathsf{RQ}[0] \odot \mathsf{DIAG}[0]$
\For{$k = 1$ to $d-1$}
  \State $\mathsf{acc} \gets \mathsf{acc} + \mathsf{RQ}[k] \odot \mathsf{DIAG}[k]$
\EndFor
\Statex \Comment{$\mathsf{acc}[i] = \langle q, v_i\rangle$ for vector $i$ of the block}
\For{$s = 1$ to $\mathit{nsq}$}
  \State $\mathsf{acc} \gets \textsc{Square}(\mathsf{acc})$
         \Comment{sharpen: raise scores to $p = 2^{\mathit{nsq}}$}
\EndFor
\For{$r = 0$ to $R-1$}
  \If{$M[r][b] = 1$}
    \State $\mathsf{rows}[r] \gets \mathsf{rows}[r] + \mathsf{acc}$
           \Comment{fold into selected rows}
  \EndIf
\EndFor
\end{algorithmic}
\end{algorithm}
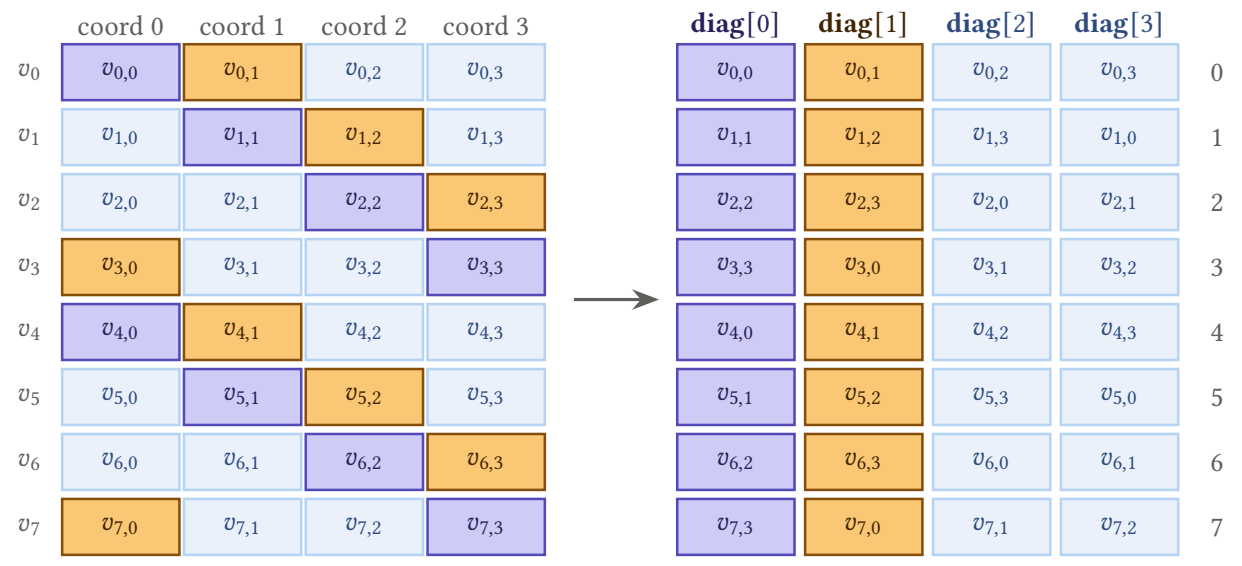
\begin{figure}[t]
  \centering
  \resizebox{\columnwidth}{!}{
%
\begingroup
\def\CarouselDiagonalFontSize{9.0} 

\pgfmathsetmacro{\CarouselDiagonalLeading}{1.2*\CarouselDiagonalFontSize}
\def\CarouselDiagonalFont{%
  \fontsize{\CarouselDiagonalFontSize pt}{\CarouselDiagonalLeading pt}\selectfont}

\definecolor{d0f}{HTML}{CECBF6}\definecolor{d0e}{HTML}{534AB7}\definecolor{d0t}{HTML}{26215C}
\definecolor{d1f}{HTML}{FAC775}\definecolor{d1e}{HTML}{854F0B}\definecolor{d1t}{HTML}{412402}
\definecolor{nf}{HTML}{EAF1FB}\definecolor{ne}{HTML}{B5D4F4}\definecolor{nt}{HTML}{2E4A7F}
\definecolor{lbl}{HTML}{5F5E5A}

\begin{tikzpicture}[
  x=1.35cm, y=0.72cm,
  font=\CarouselDiagonalFont,
  slot/.style={minimum width=1.3cm, minimum height=0.62cm,
               line width=0.8pt, font=\CarouselDiagonalFont, inner sep=0pt},
  dlbl/.style={font=\CarouselDiagonalFont\bfseries},
  hlbl/.style={text=lbl, font=\CarouselDiagonalFont},
]
\def\fillof#1{\ifcase#1 d0f\or d1f\else nf\fi}
\def\edgeof#1{\ifcase#1 d0e\or d1e\else ne\fi}
\def\textof#1{\ifcase#1 d0t\or d1t\else nt\fi}

\foreach \i in {0,...,7}{
  \foreach \j in {0,...,3}{
    \pgfmathtruncatemacro{\k}{mod(\j-\i+8,4)}
    \node[slot, draw, fill=\fillof{\k}, draw=\edgeof{\k}, text=\textof{\k}]
      (v\i\j) at (\j,-\i) {$v_{\i,\j}$};
  }
  \node[hlbl, left=0.1cm of v\i0] {$v_{\i}$};
}
\foreach \j in {0,...,3}{
  \node[hlbl] at (\j,0.72) {coord $\j$};
}

\draw[-{Stealth[length=3mm]}, line width=1pt, lbl]
  (3.72,-3.5) -- (4.42,-3.5);

\foreach \k in {0,...,3}{
  \pgfmathsetmacro{\col}{5.05+1.05*\k}
  \node[dlbl, text=\textof{\k}] at (\col,0.72) {diag$[\k]$};
  \foreach \i in {0,...,7}{
    \pgfmathtruncatemacro{\jj}{mod(\i+\k,4)}
    \node[slot, draw, fill=\fillof{\k}, draw=\edgeof{\k}, text=\textof{\k}]
      at (\col,-\i) {$v_{\i,\jj}$};
  }
}
\foreach \i in {0,...,7}{
  \node[hlbl] at (9,-\i) {$\i$};
}
\end{tikzpicture}
\endgroup}
  \caption{Diagonal plaintexts with the first two diagonals highlighted.}
  \label{fig:diag}
\end{figure}

\subsection{Encrypted Scan}
\label{sec:design:scan}

\smallskip\noindent\textbf{Query expansion.}
The scan needs one copy of the query aligned to each of the $d$ diagonals. The client
uploads only one ciphertext, so the server produces the rest itself. The uploaded query
repeats with period $d$, so rotating it by one slot advances every slot to the next
coordinate. The server therefore builds copy $k$ by rotating the uploaded query $k$
slots, giving $\mathsf{RQ}[0], \dots, \mathsf{RQ}[d-1]$. Every copy is one further
single-slot rotation of the previous one, so the whole chain only needs a single rotation
key. The expansion runs once per query and its output is reused across all $B$ blocks.
This costs memory since, the $d$ rotated copies stay resident for the duration of the scan.

\smallskip\noindent\textbf{Scoring.}
Copy $k$ and diagonal $k$ now match slot by slot. In slot $i$, copy $k$ holds
$q[(i{+}k)\bmod d]$ and diagonal $k$ holds $v_i[(i{+}k)\bmod d]$, so their product is a
matching pair of coordinates from the same vector. Multiplying the two and accumulating
over all $d$ diagonals gives one block of scores,
\begin{equation}
  \mathsf{acc} \;=\; \sum_{k=0}^{d-1} \mathsf{RQ}[k] \odot \mathsf{DIAG}[k],
  \label{eq:scan}
\end{equation}
where $\odot$ is the slot-wise product of a ciphertext against a plaintext. Slot $i$ of
$\mathsf{acc}$ now holds $\langle q, v_i \rangle$, the score of the $i$-th vector of the
block. Figure~\ref{fig:scan} shows the four products and their sum on the running
example.

This diagonal arrangement allows the system to independently accumulate the inner products in parallel. Every product landing in a given slot belongs to the same vector, so no slot ever mixes contributions from two different vectors. For instance, consider slot $0$ in Figure~\ref{fig:scan}. Across the four products it holds $q_0 v_{0,0}$, $q_1 v_{0,1}$, $q_2 v_{0,2}$ and $q_3 v_{0,3}$, which are the four terms of $\langle q, v_0\rangle$. Summing the four ciphertexts therefore places $\langle q, v_0\rangle$ in slot $0$, and the same holds for all other slots for its own vector. Since, no operation moves a value between slots, the score of vector $i$ stays in slot $i$ for the remainder of the pipeline. Since a CKKS slot holds a complex number while a score is real, we pack two blocks per ciphertext and separate them after the scan with one conjugation, halving the ciphertexts scanned at the cost of a conjugation key.

\begin{figure}[t]
  \centering
  \resizebox{\columnwidth}{!}{
%
%
\begin{tikzpicture}[x=0.62cm, y=0.74cm,
  /utils/exec={%
    \definecolor{d0f}{HTML}{CECBF6}\definecolor{d0e}{HTML}{534AB7}%
    \definecolor{d0t}{HTML}{26215C}\definecolor{d1f}{HTML}{FAC775}%
    \definecolor{d1e}{HTML}{854F0B}\definecolor{d1t}{HTML}{412402}%
    \definecolor{nf}{HTML}{EAF1FB}\definecolor{ne}{HTML}{B5D4F4}%
    \definecolor{nt}{HTML}{2E4A7F}\definecolor{qf}{HTML}{EAF3DE}%
    \definecolor{qe}{HTML}{639922}\definecolor{qt}{HTML}{27500A}%
    \definecolor{accf}{HTML}{FAECE7}\definecolor{ctedge}{HTML}{993C1D}%
    \definecolor{acct}{HTML}{712B13}\definecolor{lbl}{HTML}{5F5E5A}},
  cell/.style  ={minimum width=0.62cm, minimum height=0.64cm, inner sep=0pt,
                 line width=0.7pt, font=\scriptsize},
  q/.style     ={cell, draw=qe, fill=qf, text=qt},
  acc/.style   ={cell, draw=ctedge, fill=accf, text=acct},
  dcell/.style ={cell},
  lab/.style   ={text=lbl, font=\small},
  hdr/.style   ={text=lbl, font=\small\itshape},
  dlab/.style  ={font=\small\bfseries},
  tt/.style    ={text=lbl, font=\scriptsize\ttfamily},
  arw/.style   ={draw=lbl, line width=0.8pt, ->},
]

\def\pit{1.1}                        
\def\qL{-0.61}\def\qR{8.31}          
\def\dxo{10.6}                       
\def\dL{9.99}\def\dR{18.91}          
\def\qC{3.85}\def\dC{14.45}          
\def\hy{0.53}                        
\def\trace{2}                        

\newcommand{\stage}[7]{%
  \foreach \s in {0,...,7}{%
    \pgfmathtruncatemacro{\m}{mod(\s+#2,4)}%
    \node[q]                                at (\s*\pit,#1)         {$q_{\m}$};
    \node[dcell, fill=#4, draw=#5, text=#6] at (\dxo+\s*\pit,#1)    {$v_{\s,\m}$};}%
  \draw[draw=ctedge, line width=1.1pt, rounded corners=3pt]
    (\qL,#1-\hy) rectangle (\qR,#1+\hy);
  \draw[draw=lbl, dash pattern=on 2pt off 1.6pt, line width=0.7pt, rounded corners=3pt]
    (\dL,#1-\hy) rectangle (\dR,#1+\hy);
  \node[lab, anchor=east]           at (\qL-0.25,#1) {#3};
  \node[lab, font=\normalsize]      at (9.15,#1)     {$\odot$};
  \node[dlab, text=#6, anchor=west] at (\dR+0.25,#1) {#7};}


\node[hdr] at (\qC,2.05) {tiled query, rotated (ciphertext)};
\node[hdr] at (\dC,2.05) {diagonal of the 8 vectors (plaintext)};
\foreach \s in {0,...,7}{%
  \node[lab, font=\scriptsize] at (\s*\pit,1.05)         {\s};
  \node[lab, font=\scriptsize] at (\dxo+\s*\pit,1.05)    {\s};}
\node[lab, font=\scriptsize, anchor=east] at (\qL-0.25,1.05) {slot};

\stage{0}{0}{$k=0$}{d0f}{d0e}{d0t}{$\mathrm{diag}_0$}
\stage{-2.2}{1}{$k=1$}{d1f}{d1e}{d1t}{$\mathrm{diag}_1$}
\stage{-4.4}{2}{$k=2$}{nf}{ne}{nt}{$\mathrm{diag}_2$}
\stage{-6.6}{3}{$k=3$}{nf}{ne}{nt}{$\mathrm{diag}_3$}

\foreach \y in {0,-2.2,-4.4}{%
  \draw[arw] (\qC,\y-0.62) -- (\qC,\y-1.58);
  \node[tt, anchor=west] at (\qC+0.2,\y-1.10) {Rot(1)};}

\draw[line width=1pt, draw=lbl] (\qL,-8.00) -- (\qR,-8.00);
\node[lab, font=\large, anchor=east] at (\qL-0.25,-7.88) {$\sum$};
\draw[arw] (\qC,-8.10) -- (\qC,-8.65);

\foreach \s in {0,...,7}{\node[acc] at (\s*\pit,-9.25) {$s_{\s}$};}
\draw[draw=ctedge, line width=1.1pt, rounded corners=3pt]
  (\qL,-9.25-\hy) rectangle (\qR,-9.25+\hy);
\node[text=ctedge, font=\small, anchor=east] at (\qL-0.25,-9.25) {$\mathsf{acc}$};
\node[lab, anchor=west] at (\dL,-9.25)
  {slot $i$ holds $s_i=\langle q,\,v_i\rangle$, all $8$ inner products at once};


\end{tikzpicture}}
  \caption{The scan}
  \label{fig:scan}
\end{figure}
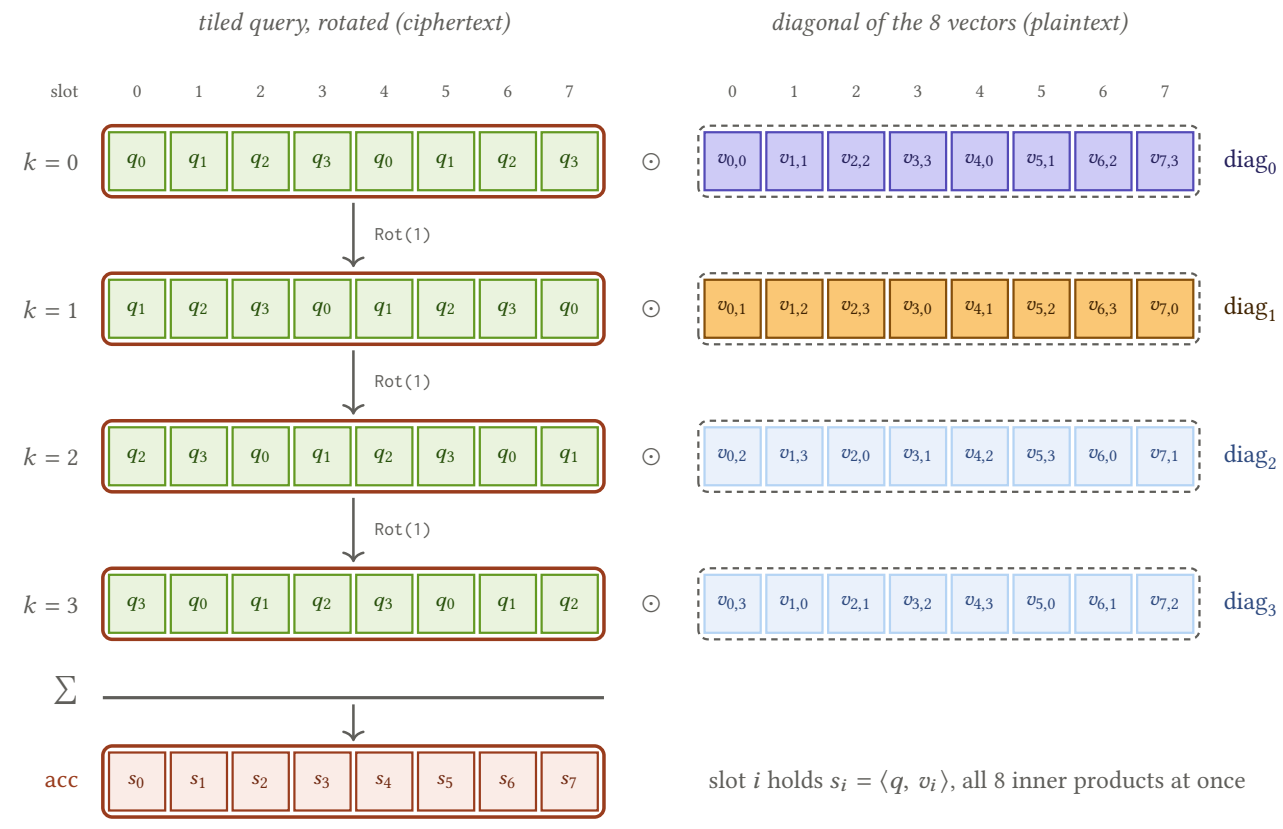

The scan operation multiplies each plaintext diagonal against the query ciphertext. However, CKKS cannot multiply a ciphertext directly by a list of numbers. The plaintext must first be converted into the number-theoretic transform (NTT) domain, a representation required for CKKS multiplication. This transform is applied once for every prime in the modulus chain. The resulting representation is much larger than the coordinates it came from. We call this \emph{diagonal encoding}. This is a computationally expensive step and dominates query latency, so the encoded diagonals are cached and reused across queries when they fit in device memory.

\subsection{Sharpening the Score Vector}
\label{sec:design:sharpen}
\label{sec:sharpen}

The vector inner products after the scan are not sparse. A large number of documents score modestly and the good matches are only slightly above them. A linear sketch cannot compress scores with this distribution. We therefore reshape it by raising every score to a power $p$, which the server does by squaring $\mathit{nsq} = \log_2 p$ times. Each squaring costs one level and one relinearisation. Squaring contracts every score, but it contracts small scores faster than large ones. A score of $0.95$ raised to the power $16$ becomes $0.44$, while a score of $0.60$ falls to $2.8 \times 10^{-4}$ and a score of $0.30$ to $4.3 \times 10^{-9}$. The gap between the head and the bulk therefore widens by orders of magnitude. After $\mathit{nsq}$ squarings only a handful of entries remain within a few percent of the peak, and the rest are numerically zero. The score vector is now sparse enough for the sketch to compress.

The operation also preserves the order of the surviving scores, so sharpening costs no
accuracy in the ranking. This holds for non-negative scores. An even power maps $-s$ and
$s$ to the same value, so a score close to $-1$ would survive sharpening and be reported
as a strong match. In practice this does not arise, because the most negative score a
query encounters is far smaller in magnitude than its largest positive score and the
ratio is suppressed by the power $p$.

The exponent depends on the distribution of the cosine scores of the corpus. A small $p$ leaves the bulk uncrushed, and the decoder cannot separate the true matches. On the other hand, a large $p$ drives even the strong scores below the noise floor, where they are lost. The usable range depends on how cleanly a corpus's true neighbors separate from the bulk, and on the corpus size.
The exponent also fixes the encryption level, $L_{\mathrm{enc}} = \mathit{nsq}+1$, so
changing $p$ changes the depth of the modulus chain.

\subsection{The Fold}
\label{sec:design:fold}
\label{sec:fold}

After sharpening, the server holds one ciphertext of scores for each of the $B$ blocks, referred to as \emph{block ciphertexts}. Only a few entries carry significant magnitude, referred to as \emph{heavy items}. The client only needs to recover the top-k scores, but sending all $B$ block ciphertexts constitutes a significant overhead. The fold compresses the $B$ ciphertexts into a smaller number of ciphertext sketches called \emph{rows}. Each row is a sum of a different subset of blocks. The goal for the client is to decrypt the rows and recover only the heavy entries.

The first row of the fold is the sum of all $B$ block ciphertexts. Slot $i$ of this row holds the sum of all scores at position $i$ across all $B$ block ciphertexts. Since only a few entries are heavy, the slots of larger magnitude in this row are the slots occupied by heavy items. The first row therefore identifies all candidate slots that the client needs to examine. Once we have identified which slots the heavy items come from, the next step is to recover which of the $B$ block ciphertexts contains the heavy item in slot $i$.

The remaining rows identify which block ciphertext contains the heavy item. A block index below $B$ requires $m=\lceil\log_2 B\rceil$ bits. To illustrate the basic construction, assign each block its $m$-bit binary codeword and create one fold row per bit position. Row $r$ sums the block ciphertexts whose $r$-th codeword bit is one, so each block is added to a unique combination of rows. These codewords form the columns of a binary \emph{fold matrix} $M$, where $M[r][b]=1$ if block ciphertext $b$ is added to row $r$. Figure~\ref{fig:fold} illustrates this uncoded construction for $B=8$, which requires three codeword rows in addition to $\mathrm{row}_0$. Block ciphertext $6$ has codeword $(0,1,1)$ (LSB first) and is therefore added to $\mathrm{row}_2$ and $\mathrm{row}_3$, but not to $\mathrm{row}_1$. A heavy item from this block appears in the corresponding pattern of rows, allowing the client to recover its block index.

In the ideal case, a heavy item occupies a given slot in exactly one block ciphertext, while every other block ciphertext holds a value near zero at that position. The client decrypts the rows and inspects $\mathrm{row}_0$ to identify the heavy slots. For each candidate slot, the client evaluates which of the remaining rows contain a large value and which remain near zero. This binary pattern directly reconstructs the codeword of the originating block ciphertext. For example, consider slot 3 in Figure~\ref{fig:fold}, where all block ciphertexts except block ciphertext 6 hold near-zero values, and block ciphertext 6 holds a magnitude of $\alpha$. Examining slot 3 across the folded rows, $\mathrm{row}_0$ will contain $\alpha$, flagging it as a slot of interest. Extracting the values across all rows at this slot yields the vector $\langle \alpha, 0, \alpha, \alpha \rangle$. By matching this non-zero pattern to the known codewords, the client determines that the item originated from block ciphertext 6.

\begin{figure}[t]
  \centering
  \resizebox{\columnwidth}{!}{
\begingroup
\def\CarouselFoldFontSize{12} 

\pgfmathsetmacro{\CarouselFoldLeading}{1.2*\CarouselFoldFontSize}
\pgfmathsetmacro{\CarouselFoldHeavySize}{1.2*\CarouselFoldFontSize}
\pgfmathsetmacro{\CarouselFoldHeavyLeading}{1.2*\CarouselFoldHeavySize}
\def\CarouselFoldFont{%
  \fontsize{\CarouselFoldFontSize pt}{\CarouselFoldLeading pt}\selectfont}
\def\CarouselFoldHeavyFont{%
  \fontsize{\CarouselFoldHeavySize pt}{\CarouselFoldHeavyLeading pt}\selectfont}

\begin{tikzpicture}[x=0.62cm, y=0.74cm,
  /utils/exec={%
    \definecolor{accf}{HTML}{FAECE7}\definecolor{ctedge}{HTML}{993C1D}%
    \definecolor{acct}{HTML}{712B13}\definecolor{nf}{HTML}{EAF1FB}%
    \definecolor{ne}{HTML}{B5D4F4}\definecolor{nt}{HTML}{2E4A7F}%
    \definecolor{lbl}{HTML}{5F5E5A}},
  slot/.style  ={minimum width=0.56cm, minimum height=0.60cm, inner sep=0pt,
                 line width=0.7pt, rounded corners=1.5pt,
                 draw=ctedge!55, fill=accf},
  heavy/.style ={slot, fill=ctedge!45, draw=ctedge, text=acct,
                font=\CarouselFoldHeavyFont\boldmath},
  tcell/.style ={minimum width=0.70cm, minimum height=0.74cm, inner sep=0pt,
                 line width=0.5pt, draw=lbl!55, fill=white,
                 font=\CarouselFoldFont, text=lbl},
  lab/.style   ={text=lbl, font=\CarouselFoldFont},
  hdr/.style   ={text=lbl, font=\CarouselFoldFont\itshape},
  tiny/.style  ={text=lbl, font=\CarouselFoldFont},
  flow/.style  ={draw=ctedge, line width=0.9pt,
                 -{Stealth[length=4.5pt,width=3.6pt]}},
  skip/.style  ={draw=lbl!70, line width=0.8pt, dash pattern=on 2.4pt off 2pt,
                 -{Stealth[length=4pt,width=3.2pt]}},
]

\def\tx{8.65}
\def\tp{1.15}
\def\rx{14.70}
\def\hv{3}
\def\selfill#1{\ifcase#1 white\or nf\fi}
\def\seltext#1{\ifcase#1 lbl\or ctedge\fi}
\def\selrowt#1{\ifcase#1 lbl!70\or acct\fi}

\foreach \s in {0,...,7}{\node[tiny] at (\s,1.05) {\s};}
\node[tiny, anchor=east] at (-0.75,1.05) {slot};
\foreach \i in {0,...,7}{%
  \ifnum\i=6\relax
    \node[lab, anchor=east, text=ctedge,
          font=\CarouselFoldFont\bfseries\boldmath]
      at (-0.75,-\i) {$\mathsf{ct}_{\i}$};
  \else
    \node[lab, anchor=east] at (-0.75,-\i) {$\mathsf{ct}_{\i}$};
  \fi
  \foreach \s in {0,...,7}{\node[slot] at (\s,-\i) {};}}
\foreach \s in \hv{\node[heavy] at (\s,-6) {$\alpha$};}

\node[tcell, fill=lbl!10, draw=lbl!55] at (\tx,1.05) {$b$};
\foreach \c in {1,2,3}{%
  \node[tcell, fill=lbl!10, draw=lbl!55]
    at (\tx+\c*\tp,1.05) {$c_{\c}$};}
\foreach \b in {0,...,7}{%
  \pgfmathtruncatemacro{\hit}{ifthenelse(\b==6,1,0)}%
  \node[tcell, fill=\selfill{\hit}] at (\tx,-\b) {$\b$};
  \foreach \c in {1,2,3}{%
    \pgfmathtruncatemacro{\bit}{mod(int(\b/(2^(\c-1))),2)}%
    \node[tcell, fill=\selfill{\hit}, text=\seltext{\hit}]
      at (\tx+\c*\tp,-\b) {$\bit$};}}
\draw[draw=ctedge, line width=1pt, rounded corners=2pt]
  (\tx-0.68,-6.60) rectangle (\tx+3*\tp+0.68,-5.40);

\foreach \s in {0,...,7}{\node[tiny] at (\rx+\s,-0.05) {\s};}
\newcommand{\rrow}[4]{%
  \foreach \s in {0,...,7}{\node[slot] at (\rx+\s,#1) {};}%
  \ifcase#4\relax\or
    \foreach \s in \hv{\node[heavy] at (\rx+\s,#1) {$\alpha$};}%
  \fi
  \node[anchor=west, font=\CarouselFoldFont, text=\selrowt{#4}]
    at (\rx-0.45,#1-0.72) {$\mathsf{row}_{#2}$\;\textemdash\;#3};}
\rrow{-0.9}{0}{all ones}{1}
\rrow{-2.7}{1}{$c_1=0$}{0}
\rrow{-4.5}{2}{$c_2=1$}{1}
\rrow{-6.3}{3}{$c_3=1$}{1}

\draw[flow] (\tx+3*\tp+0.85,-5.75)
  to[out=62,in=182] (\rx-0.55,-0.9);
\draw[skip] (\tx+3*\tp+0.85,-6.00) to[out=40,in=186]
  node[pos=0.72, fill=white, inner sep=1.2pt, text=lbl!80,
       font=\CarouselFoldFont\bfseries\boldmath] {$\times$}
  (\rx-0.55,-2.7);
\draw[flow] (\tx+3*\tp+0.85,-6.15)
  to[out=-12,in=190] (\rx-0.55,-4.5);
\draw[flow] (\tx+3*\tp+0.85,-6.35)
  to[out=-46,in=194] (\rx-0.55,-6.3);


\end{tikzpicture}
\endgroup}
  \caption{Folding $B=8$ blocks into four rows.}
  \label{fig:fold}
\end{figure}
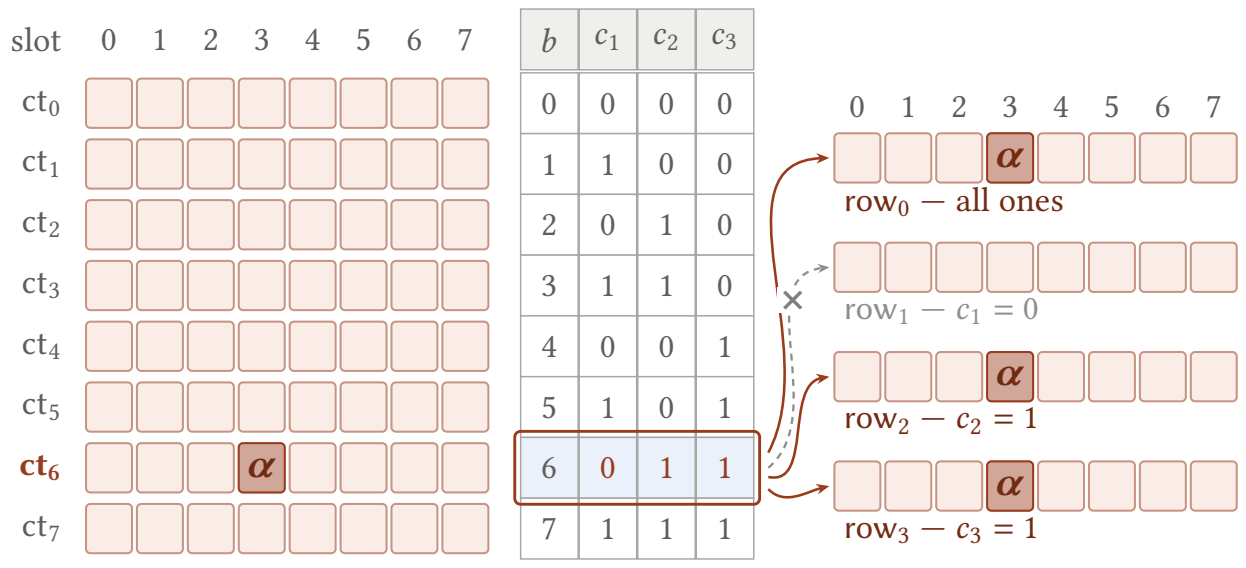

The idealised assumption fails due to two reasons. First, the block ciphertexts that do not hold the heavy item do not contribute zero. Each row sums many block ciphertexts, and every block carries a residual bulk left over from sharpening, so a slot that should contain a near zero value instead holds the accumulated bulk of all the block ciphertexts added into that row, together with the encryption noise of CKKS. Therefore the entries are not exact but large and small values, and the more block ciphertexts contribute to a row, the larger the small values grow. Reading each row as an exact bit becomes unreliable due to noise corrupting some of the values. Second, two heavy items can occupy the same slot in two different block ciphertexts. Their codewords are then added together, and the client sees the sum of two codewords rather than one. As a result, the collision of two heavy items in the same slot, introduces a secondary challenge for the client-side decoding phase.

The first problem is resolved by replacing standard binary labels with a more robust codeword assignment. With plain binary codewords, two block codewords can differ in a single bit, for example, codewords of blocks ciphertext $4$ and $5$ differ only in the last position. This means a single unreliable row can turn one block's pattern into another's. The solution is to read the rows not as exact bits, but as a pattern to be matched to the nearest codeword. To ensure that this nearest match remains unambiguous under noise, we draw the codewords from a code of minimum Hamming distance three\cite{kautz}. Because any two codewords now differ in at least three positions, the true codeword stays closest to the observed pattern even if one row is completely corrupted. This spacing makes the decode robust to the residual bulk and encryption noise.

\textsc{Carousel} constructs the robust codewords using a randomized procedure. It begins with the $m$-bit representation of each block ID and repeatedly samples a random linear combination of the ID bits, appending the resulting bit to every block's codeword. Sampling continues until every pair of codewords has Hamming distance at least $3$, producing an $L$-bit base codeword. \textsc{Carousel} then samples $r_p$ additional combinations in the same manner and appends them as extra measurement bits, yielding a codeword $c(b)$ of length $L+r_p$. Each codeword position defines one fold row, so together with the all-ones $\mathrm{row}_0$, the server returns $ R = 1 + L + r_p $ rows. The augmented codewords form the columns of $M\in{0,1}^{(R-1)\times B}$. Because appending coordinates cannot decrease Hamming distance, the augmented code retains a minimum distance of at least $3$.

At any identified candidate slot, the client extracts the corresponding values from each row to construct an observation vector $y$. This vector is composed of three distinct contributions. First, a heavy item of magnitude $\alpha$ residing in block ciphertext $b$ contributes $\alpha$ exclusively to the rows where its assigned codeword $c(b)$ contains a one, generating a structured component $\alpha\, c(b)$. Second, because every code row aggregates the residual bulk from its constituent block ciphertexts, this accumulation introduces a uniform baseline offset $\beta$ across all rows, represented by the term $\beta\, \mathbf{1}$. Finally, the inherent CKKS encryption noise and minor inter-row variations in the bulk are absorbed into a single error vector $\varepsilon$. The observed vector is therefore formalized as
\begin{equation}
  y \;=\; \alpha\, c(b) \;+\; \beta\, \mathbf{1} \;+\; \varepsilon,
  \label{eq:slotmodel}
\end{equation}
which describes a scaled codeword shifted by a constant offset and perturbed by noise. Recovering the originating block ciphertext thus reduces to determining which valid codeword $c(b)$ best aligns with this observation signal.

To determine which codeword best aligns with the observation, we treat the evaluation of each candidate as a two-parameter fit. Assuming the heavy item originated in a specific block ciphertext $b$, the vector $y$ should approximate $\alpha\, c(b) + \beta\, \mathbf{1}$ for some magnitude $\alpha$ and offset $\beta$, disregarding the error term $\varepsilon$. For a given codeword, the optimal fit is the pair $(\alpha, \beta)$ that minimizes the distance to $y$. The quality of this fit is measured by the residual:
\begin{equation}
  \mathrm{res}(b) \;=\; \min_{\alpha,\, \beta}\;
  \big\lVert\, y - \alpha\, c(b) - \beta\, \mathbf{1} \,\big\rVert^{2}.
  \label{eq:residual}
\end{equation}
This residual is small only when $c(b)$ can be scaled and shifted to reproduce the observed pattern. The client therefore evaluates $\mathrm{res}(b)$ for every block ciphertext and selects the candidate with the minimum residual. The fitted magnitude $\alpha$ at that optimal block then provides an estimate of the item's score, allowing the client to rank the recovered matches. The block index from the residual fit and the slot index from the candidate position together give the item's global identifier, $\mathrm{id} = b \cdot \ell + i$. This is repeated over the candidate slots and ordering by the fitted magnitude yields the top matches.

The second problem, a collision, is when two heavy items occupy the same slot in different block ciphertexts. The observation by the client is a superposition, $y \approx \alpha_1 c(b_1) + \alpha_2 c(b_2) + \beta\mathbf{1} + \varepsilon$, and the single-codeword fit of Equation~\ref{eq:residual} can match at most one of them. A collision is a birthday event over the $\ell$ slots of a block ciphertext. It depends on the number of surviving heavy items and the slot count, and not on $N$ or $B$, so it does not become more frequent as the corpus grows. The rate also remains low since $k \ll \ell$ items survive sharpening. Among the top hundred scores, the expected number of colliding pairs in our settings is 0.15 per query, so most queries have no collision at all and the few that do, result in at most one match. We recover the second item by a second pass over the residual. Having fitted the first item's magnitude $\alpha_1$ at a slot, the client subtracts its contribution $\alpha_1 c(b_1)$ from the observation and refits the remainder against every codeword, excluding the block ciphertext already chosen. Subtracting the first item makes the second fit informative since, any correlation that survives is evidence of a genuinely different source. A second item is accepted only when its fitted magnitude is significant compared to the first. 

\begin{algorithm}[t]
\caption{Client decoding of one query}
\label{alg:decode}
\begin{algorithmic}[1]
\Require decrypted rows $C_0 \ldots C_{R-1}$, fold matrix $M$, budget $C$
\State $\mathit{cand} \gets$ indices of the $C$ largest entries of $C_0$
\ForAll{$i \in \mathit{cand}$}
  \State $y \gets (C_1[i], \ldots, C_{R-1}[i])$
  \State $\tilde y \gets y - \mathrm{mean}(y)$
  \State $\hat b_i \gets \arg\max_b
         \langle \tilde y, \tilde c(b)\rangle^{2} / \lVert \tilde c(b)\rVert^{2}$
  \State $\hat s_i \gets
         \langle \tilde y, \tilde c(\hat b_i)\rangle / \lVert \tilde c(\hat b_i)\rVert^{2}$
\EndFor
\State \Return $\{\, \hat b_i \cdot \ell + i \,\}$ ordered by decreasing $\hat s_i$
\end{algorithmic}
\end{algorithm}

The residual fit of Equation~\ref{eq:residual} has a closed form that the client can evaluate. In order to minimize the residue over the offset $\beta$, we subtract the mean from both $y$ and the codeword, which removes the constant bulk term $\beta\mathbf{1}$ and leaves only the block-dependent part. Minimising over the magnitude $\alpha$ is then a projection of the centred observation onto the centred codeword. Therefore, the block index with smallest residual is the one with largest normalised correlation.

\begin{equation}
  \hat b \;=\; \arg\min_b \mathrm{res}(b)
           \;=\; \arg\max_b
           \frac{\langle \tilde y, \tilde c(b)\rangle^{2}}{\lVert \tilde c(b)\rVert^{2}},
  \label{eq:decode}
\end{equation}
where $\tilde y$ and $\tilde c(b)$ are the mean-centred observation and codeword. The fitted magnitude $\hat s = \langle \tilde y, \tilde c(\hat b)\rangle / \lVert \tilde c(\hat b)\rVert^{2}$ estimates the item's score. This is a matched filter, correlating the observation against each candidate codeword
and selecting the largest is the optimal linear detector when a single codeword is present~\cite{aldridge}. The $r_p$ additional rows give the matched filter more measurements, improving its robustness to residual bulk and CKKS noise.

In Algorithm~\ref{alg:decode}, we show the client side decode. It selects candidate slots from $C_0$, and at each one recovers the block index using Equation~\ref{eq:decode} and its score by the estimated magnitude ($\alpha$). The collision pass (excluded in algorithm ~\ref{alg:decode}) extends this without any changes. After the first block index is fitted, the client subtracts $\hat s\,\tilde c(\hat b)$ from $\tilde y$ and applies the same Equation~\ref{eq:decode} to the residual, excluding the block index already chosen.

\subsection{Updates}
\label{sec:design:updates}

\textsc{Carousel} maintains a flat index which is a array of vectors divided into many blocks. These blocks are encoded into diagonals and each query reads from these encoded diagonals. Each group of $d$ diagonals contain coordinates from $\ell$ vectors of the same block. Since, a vector contributes one entry to each of its block's $d$ diagonals, the cost of an update is re-encoding of these $d$ diagonals. 

\smallskip\noindent\textbf{Insert.}
An insert appends the new vector to the array. If the last block has available rows, the vector fills the first spare row. Otherwise, a new block is allocated for it. Because the vector index and the downstream PIR store share a single identifier space, they must remain synchronized. The record is appended to the PIR store first, and then the new vector is added to the vector index. This ordering guarantees that every identifier returned by the search is already present in the PIR. Consequently, the underlying PIR scheme must support append operations~\cite{checklist,singlepass}. If the block count $B$ increases  after any insertion, the fold matrix $M$ and the number of rows $R$ also need to be recomputed.  Since $M$ is public and entirely generated by a shared seed, the server transmits only the new $B$ and each client regenerates $M$ locally.

\smallskip\noindent\textbf{Delete.}
A delete zeroes all of the $d$ coordinates of the target vector. A zero vector has inner product zero with every query, so the vector scores zero before and after sharpening, and has no effect on the fold row. Thus, once the block is re-encoded the index of the vector is unreachable. Because the search never exposes the identifier again, the matching PIR record does not need to be removed immediately. It can be lazily reclaimed later during compaction.

\section{Evaluation}
\label{sec:eval}

In this section, we evaluate \textsc{Carousel}’s retrieval quality, efficiency, scalability, and update performance. We also conduct various ablation studies to examine its principal tradeoffs, including projection width versus retrieval quality and response size versus recall. Finally, we compare \textsc{Carousel} with prior private vector-search systems.

\smallskip\noindent\textbf{Experimental setup.} We run all server-side computations on a single NVIDIA RTX 5090 GPU with 32GB of device memory under CUDA 13.2, hosted on a machine with 96GB of RAM, unless otherwise noted. Homomorphic evaluation uses Cheddar~\cite{cheddar}, a CKKS library using 32-bit words, with ring degree $65,536$ and scaling factor $2^{30}$. We measure all client-side latencies using a single core of an AMD Ryzen 9950X3D processor.

\smallskip\noindent\textbf{Datasets and workloads.}
We evaluate \textsc{Carousel} on four workloads. SIFT-100M~\cite{jegou2011searching} contains $10^8$ 128-dimensional vectors, which is searched at native width. For the multi-GPU evaluation, we use the full BIGANN SIFT-1B dataset~\cite{jegou2011searching}, containing $10^9$ vectors of the same dimensionality. MS MARCO~\cite{bajaj2016msmarco} contains $3.21 \times 10^6$ 768-dimensional vectors, which we project to 128 and 256 dimensions using PCA. GIST1M~\cite{jegou2011product} contains $10^6$ 960-dimensional vectors, which we project to 256 and 512 dimensions.

\smallskip\noindent\textbf{Evaluation metrics.} Each configuration is evaluated on 1,000 queries and we report recall@$k$, the fraction of the exact top-$k$ neighbors appearing in the returned set. For MS MARCO, we additionally report MRR@100, following prior private search systems~\cite{tiptoe,wally,pacmann}. A query receives a score of $1/j$ if its ground-truth passage appears at rank $j \leq 100$, and zero otherwise. MRR@100 is the mean of these scores across all queries. Because \textsc{Carousel} searches over projected vectors, we distinguish three metrics to separate the sources of retrieval loss. The \emph{projection ceiling} compares the exact top-$k$ neighbors in the projected space with those in the original space, measuring the quality retained after dimensionality reduction alone. \emph{Projected recall} compares \textsc{Carousel}'s output with the exact top-$k$ in the projected space, isolating loss introduced by the search pipeline. Finally, \emph{true recall} compares its output with the exact top-$k$ in the original, unprojected space, capturing the combined end-to-end loss from projection and the encrypted pipeline.

\smallskip\noindent\textbf{Security of Deployed Parameters. }The deployed instance uses ring degree $2^{16}$, a ternary secret of Hamming weight $1024$, Gaussian error of standard deviation $3.2$, and evaluation keys modulo a product reaching $2^{610}$. Under the lattice estimator, the cheapest known attack on this configuration costs $2^{391}$, above the $256$-bit category of the homomorphic encryption standard~\cite{hestandard,estimator}.

\subsection{Retrieval Quality}

Table~\ref{tab:quality} reports \textsc{Carousel}'s retrieval quality on the three datasets, using the native dimensionality for SIFT and and two projected dimensionalities each for MS MARCO and GIST. Projected recall ranges from $0.9411$ to $0.9887$, showing that the encrypted pipeline recovers at least $94\%$ of the neighbors returned by an exact search in the projected space. Moreover, true recall remains close to the corresponding projection ceiling, indicating that dimensionality reduction, rather than the encrypted pipeline, accounts for most of the end-to-end quality loss.

\begin{table}[t]
\centering\small
\caption{Retrieval quality over 1,000 queries per configuration.}
\label{tab:quality}
\resizebox{\columnwidth}{!}{%
\begin{tabular}{lrrrrr}
\toprule
& SIFT & \multicolumn{2}{c}{MS MARCO} & \multicolumn{2}{c}{GIST} \\
\cmidrule(lr){2-2}\cmidrule(lr){3-4}\cmidrule(lr){5-6}
dim.\ $\to d$
    & $128$
    & $768{\to}128$
    & $768{\to}256$
    & $960{\to}256$
    & $960{\to}512$ \\
\midrule
projection ceiling, r@100
    & --
    & $0.6367$
    & $0.8639$
    & $0.8045$
    & $0.9240$ \\
projected r@100
    & --
    & $0.9797$
    & $0.9887$
    & $0.9411$
    & $0.9413$ \\
true r@100
    & $0.9845$
    & $0.6337$
    & $0.8610$
    & $0.7903$
    & $0.8996$ \\
true r@10
    & $0.9785$
    & $0.6885$
    & $0.8740$
    & $0.7531$
    & $0.8842$ \\
    

MRR@100
    & --
    & $0.2763$
    & $0.3031$
    & --
    & -- \\
\bottomrule
\end{tabular}}
\end{table}

On SIFT, \textsc{Carousel} searches the native $128$-dimensional vectors and achieves true recall of $0.9845$ at rank $100$ and $0.9785$ at rank $10$. On MS MARCO, increasing the projection width from $128$ to $256$ raises true recall@100 from $0.6337$ to $0.8610$ and MRR@100 from $0.2763$ to $0.3031$. Exact search over the full $768$-dimensional vectors achieves an MRR@100 of $0.3125$, so Carousel reaches $97.0\%$ of this plaintext ceiling.

GIST exhibits the same projection-width tradeoff. Increasing the projected dimension from $256$ to $512$ improves true recall@100 from $0.7903$ to $0.8996$ and true recall@10 from $0.7531$ to $0.8842$. Meanwhile, projected recall remains essentially unchanged at approximately $0.94$. Thus, the wider projection improves the information retained from the original vectors without changing the accuracy of the encrypted pipeline.

\subsection{Query Performance and Scalability}
\label{sec:eval:scaling}

Table~\ref{tab:performance} reports the average per-query cost across $1000$ queries for each evaluated configuration. \textsc{Carousel} supports two execution modes depending on whether the encoded database diagonals fit in device memory. In the \emph{cached mode}, the server retains these diagonals and reuses them across queries, removing diagonal encoding from the query path. In the \emph{recompute mode}, the server retains only the flat vector array and reconstructs the diagonal encodings for every query. The choice between these modes is determined by device-memory capacity. MS MARCO and GIST use cached diagonal encodings, achieving server latencies of $0.031$--$0.045$ seconds and $0.034$--$0.057$ seconds, respectively. On MS MARCO, server latency is dominated by the encrypted scan and sharpening stages. However on GIST, query expansion accounts for approximately half of the total server latency. Query expansion scales with vector dimensionality, whereas scanning and sharpening scale with dataset size. Because GIST uses equal or higher projected dimensions but contains fewer vectors than MS MARCO, query expansion constitutes a larger fraction of its latency.

SIFT represents the capacity-limited mode. Since, its encoded diagonals require approximately $512$~GB and do not fit on our $32$~GB GPU, \textsc{Carousel} recomputes the diagonal encodings for each query. Of its $3.735$-second server latency, encoding accounts for $1.993$ seconds or $53.4\%$, followed by sharpening at $1.083$ seconds or $29.0\%$ and scanning at $0.553$ seconds or $14.8\%$. Folding adds only $0.097$ seconds because it requires neither ciphertext rotations nor multiplications.

\begin{table}[h]
\centering\small
\caption{Per-query cost averaged over 1000 queries.}
\label{tab:performance}
\resizebox{\columnwidth}{!}{%
\begin{tabular}{lrrrrr}
\toprule
& SIFT & \multicolumn{2}{c}{MS MARCO} & \multicolumn{2}{c}{GIST} \\
\cmidrule(lr){2-2}\cmidrule(lr){3-4}\cmidrule(lr){5-6}
vectors $N$           & $10^{8}$ & $3.21$M & $3.21$M & $1$M & $1$M \\
dim.\ $\to d$         & $128$ & $768{\to}128$ & $768{\to}256$
                      & $960{\to}256$ & $960{\to}512$ \\
blocks $B$            & $3052$ & $99$ & $99$ & $31$ & $31$ \\
cached diagonals (GB) & $512$ & $10.07$ & $20.13$ & $9.66$ & $19.33$ \\
sharpening exponent ($p$) & $128$ & $8$ & $8$ & $64$ & $64$ \\
\midrule
encode                & $1.993$ & $--$ & $--$ & $--$ & $--$ \\
query expansion       & $0.0078$ & $0.0063$ & $0.0125$ & $0.0151$ & $0.0303$ \\
scan                  & $0.553$ & $0.0116$ & $0.0191$ & $0.0083$ & $0.0162$ \\
sharpening                 & $1.083$ & $0.0097$ & $0.0096$ & $0.0089$ & $0.0089$ \\
fold                  & $0.097$ & $0.0027$ & $0.0027$ & $0.0007$ & $0.0007$ \\
finalize              & $0.0014$ & $0.0010$ & $0.0010$ & $0.0009$ & $0.0009$ \\
\textbf{server total (sec)} & $\mathbf{3.735}$ & $\mathbf{0.031}$ & $\mathbf{0.045}$
                      & $\mathbf{0.034}$ & $\mathbf{0.057}$ \\
\midrule
client query build (ms)      & $23.31$ & $19.09$ & $18.91$ & $21.77$ & $22.10$ \\
client decrypt \& decode (ms)& $12.01$ & $9.78$ & $9.55$ & $8.50$ & $8.56$ \\
upload / download (MB)       & $5.24$/$12.58$ & $3.15$/$10.49$ & $3.15$/$10.49$
                             & $4.72$/$8.39$ & $4.72$/$8.39$ \\
eval.\ keys (MB)             & $31.46$ & $18.88$ & $18.88$ & $28.31$ & $28.31$ \\
\bottomrule
\end{tabular}}
\end{table}

Client-side computation remains efficient in all of the configurations. Building the query takes $18.9$--$23.3$ ms and decrypting and decoding the response takes $8.5$--$12.0$ ms across the five configurations. Queries upload $3.2$--$5.2$ MB and download $8.4$--$12.6$ MB. The upload size varies because the corpus-specific sharpening exponent determines the required multiplicative depth, which in turn determines the ciphertext modulus and size.

\smallskip\noindent\textbf{Effect of caching.}
To isolate the benefit of caching, Figure~\ref{fig:cache-vs-recomp} evaluates both execution modes on three configurations whose encoded diagonals fit in device memory. Caching eliminates the encoding stage while leaving query expansion, scanning, sharpening, and folding unchanged. It reduces server latency from $46.8$~ms to $26.2$~ms on SIFT-1M (the first one million vectors of SIFT), from $211.7$~ms to $44.8$~ms on MS MARCO, and from $90.3$~ms to $33.8$~ms on GIST. These reductions correspond to speedups of $1.79\times$, $4.73\times$, and $2.67\times$, respectively. The improvement is largest on MS MARCO because it uses a sharpening exponent of only $p=8$, compared with $p=128$ on SIFT-1M and $p=64$ on GIST. Since, sharpening requires less compute, encoding accounts for $79\%$ of the recomputed latency on MS MARCO, compared with $62\%$ on GIST and $45\%$ on SIFT-1M. Caching eliminates this stage and thus removes the corresponding share of the latency. The cached representations occupy $5.37$ GB, $20.13$ GB, and $9.66$~GB of device memory, respectively.

\begin{figure}[h]
\centering
\includegraphics[width=\columnwidth]{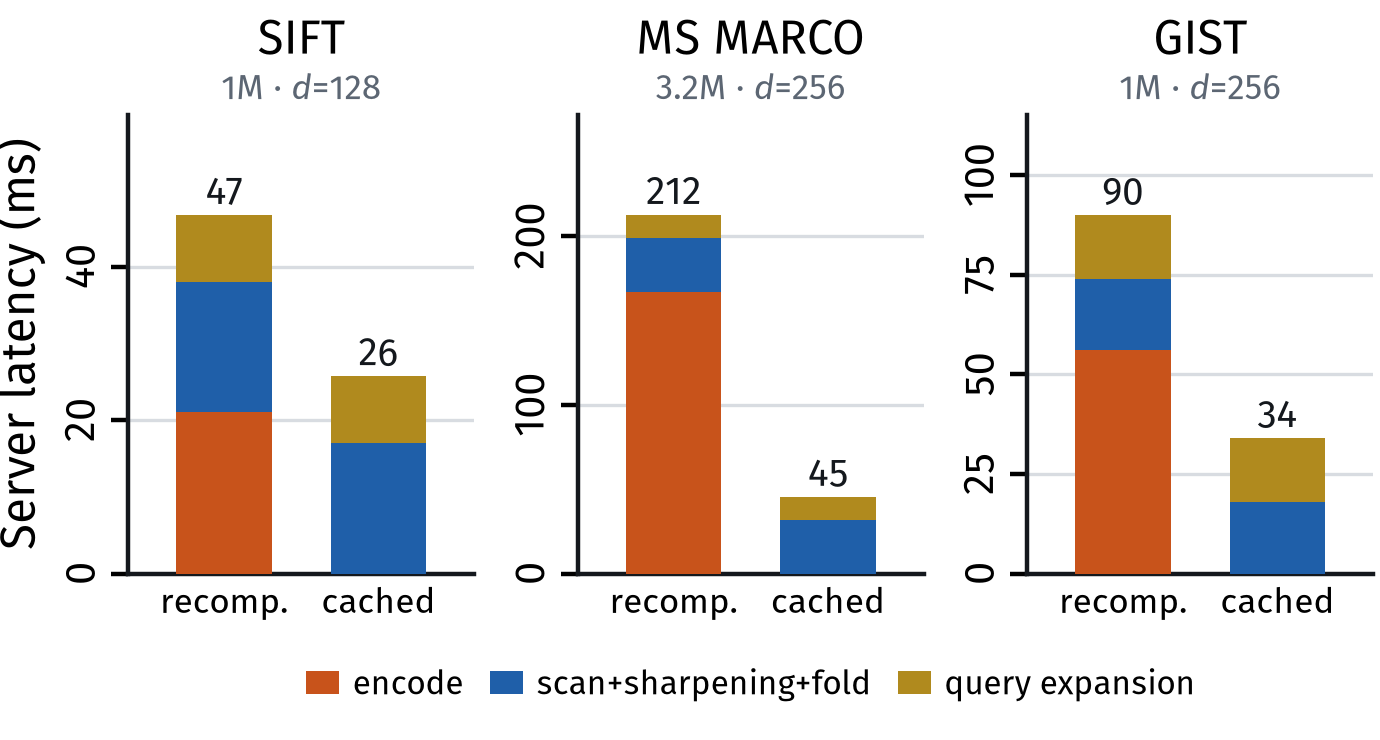}
\caption{Server latency with recomputed and cached diagonal encodings.}
\label{fig:cache-vs-recomp}
\end{figure}

\smallskip\noindent\textbf{Scaling corpus size($N$).}
We evaluate how \textsc{Carousel} scales with the corpus size $N$. Figures ~\ref{fig:marco-scaling} and ~\ref{fig:sift-scaling} report server latency and recall@$100$ as we vary $N$ on MS MARCO (cached mode) and SIFT (recompute mode). 

\begin{figure}[b]
    \centering
    \begin{subfigure}[t]{0.49\columnwidth}
        \centering
        \includegraphics[width=\linewidth]{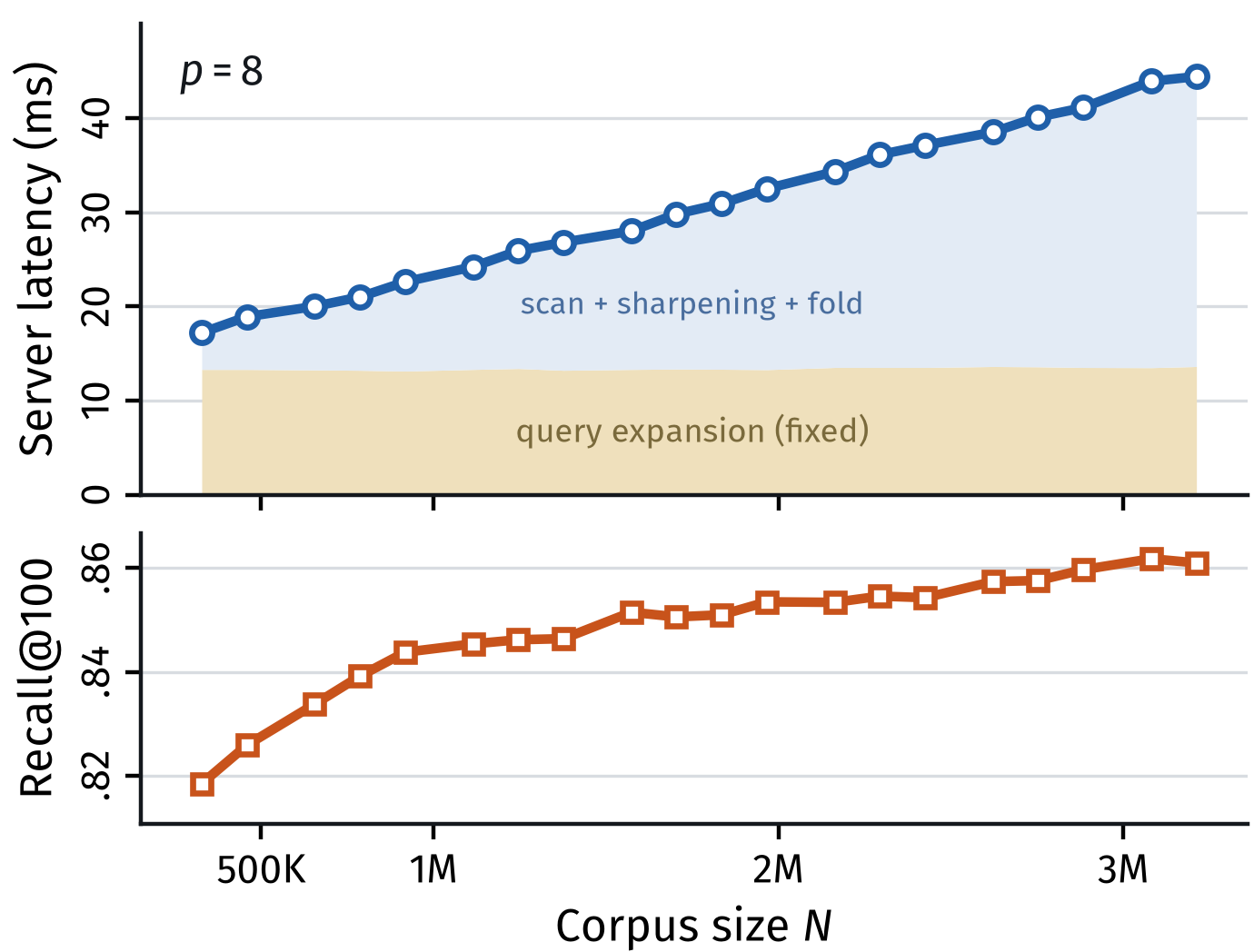}
        \caption{MS MARCO ($d=256$)}
        \label{fig:marco-scaling}
    \end{subfigure}
    \hfill
    \begin{subfigure}[t]{0.49\columnwidth}
        \centering
        \includegraphics[width=\linewidth]{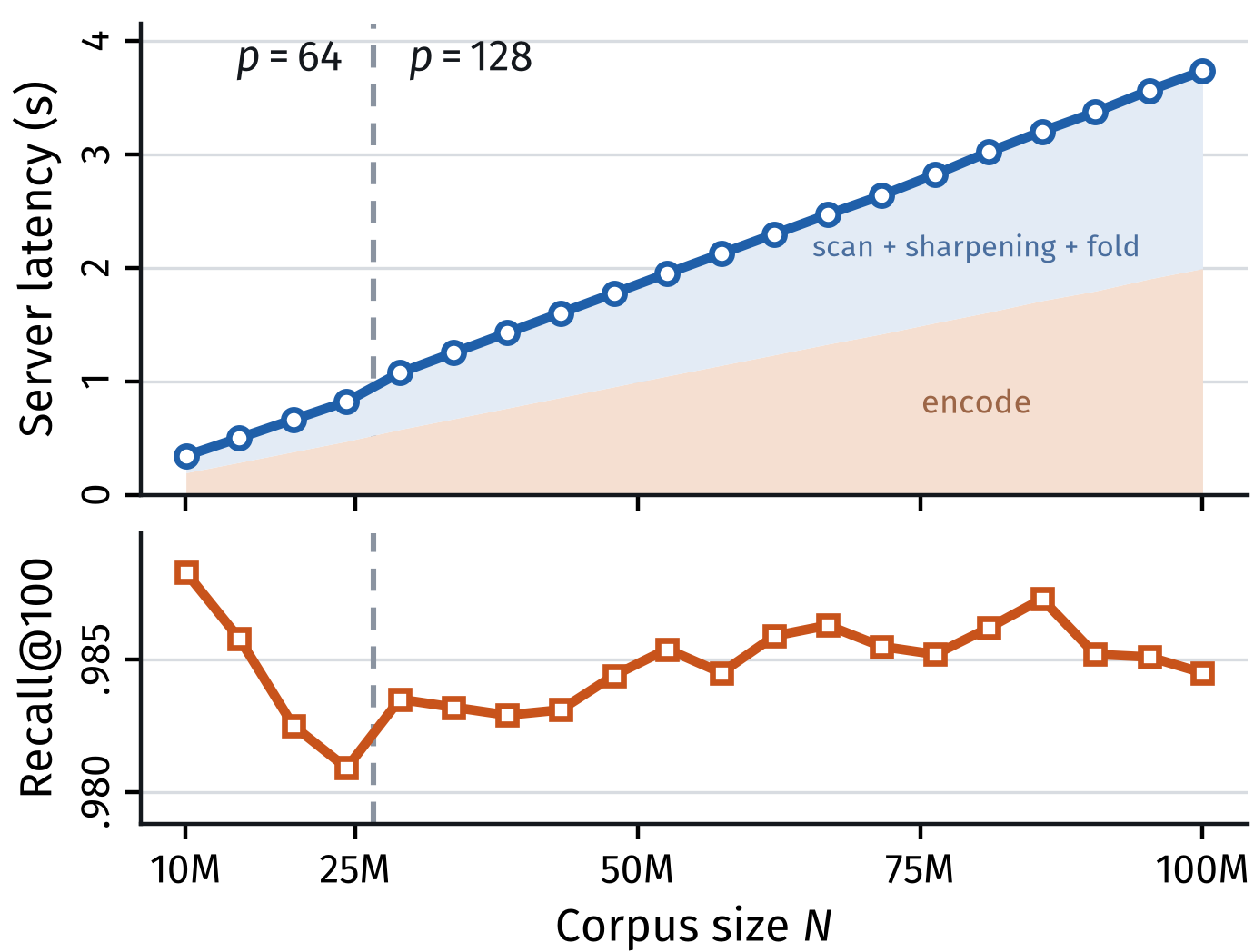}
        \caption{SIFT-100M ($d=128$)}
        \label{fig:sift-scaling}
    \end{subfigure}
    \caption{Scaling with corpus size. Server latency grows linearly with $N$ on both datasets. Recall@$100$ increases with $N$ on MS MARCO and remains stable on SIFT-100M. The dashed line marks the change from $p=64$ to $p=128$.}
    \label{fig:scaling}
\end{figure}

On SIFT, server latency grows linearly from $0.343$~s at $10$M vectors to $3.735$~s at $100$M. The decompositions shows that both the encoding stage and the combined scan--sharpening--fold stage scale proportionally with $N$, resulting in a linear growth in end-to-end server latency. We observe that, recall depends on the sharpening exponent. At $p=64$, recall declines from $0.9883$ to $0.9809$ as the corpus grows to $24.2$M vectors, since the accumulated residual bulk exceeds what this exponent can suppress. We therefore increase the exponent to $p=128$ at $29.0$M vectors, as marked by the dashed line. Recall subsequently recovers and remains between $0.9829$ and $0.9873$ as we scale further towards $100$M vectors. 

MS MARCO exhibits the same linear scaling at a fixed exponent of $p=8$. With diagonal encodings cached, increasing the corpus from $0.328$M to $3.21$M vectors raises server latency from $17.2$~ms to $44.4$~ms and recall from $0.8185$ to $0.8610$. The server latency consists of a fixed component and a component that scales with the corpus: query expansion remains constant at $12.4$--$12.6$~ms, whereas the scan--sharpening--fold stage grows with $N$. Because encoding is removed from the query path and query expansion does not scale with $N$, a $9.80\times$ increase in corpus size raises latency by only $2.58\times$.

These results confirm the expected behavior of an exhaustive encrypted scan that is server latency scales linearly with $N$, while retrieval quality remains stable.

\subsection{Response Size and Recall}
\label{sec:eval:response_size}
Recall that Section~\ref{sec:design:fold} constructs a distance-$3$ base code using randomized parity rows and augments it with $r_p$ additional measurement rows. Although every generated code satisfies the same minimum-distance guarantee, its sampled structure can affect the matched-filter decoder. We therefore evaluate the variance across generated codes and the effect of $r_p$ on retrieval quality.

\begin{figure}[h]
\centering
\includegraphics[width=\columnwidth]{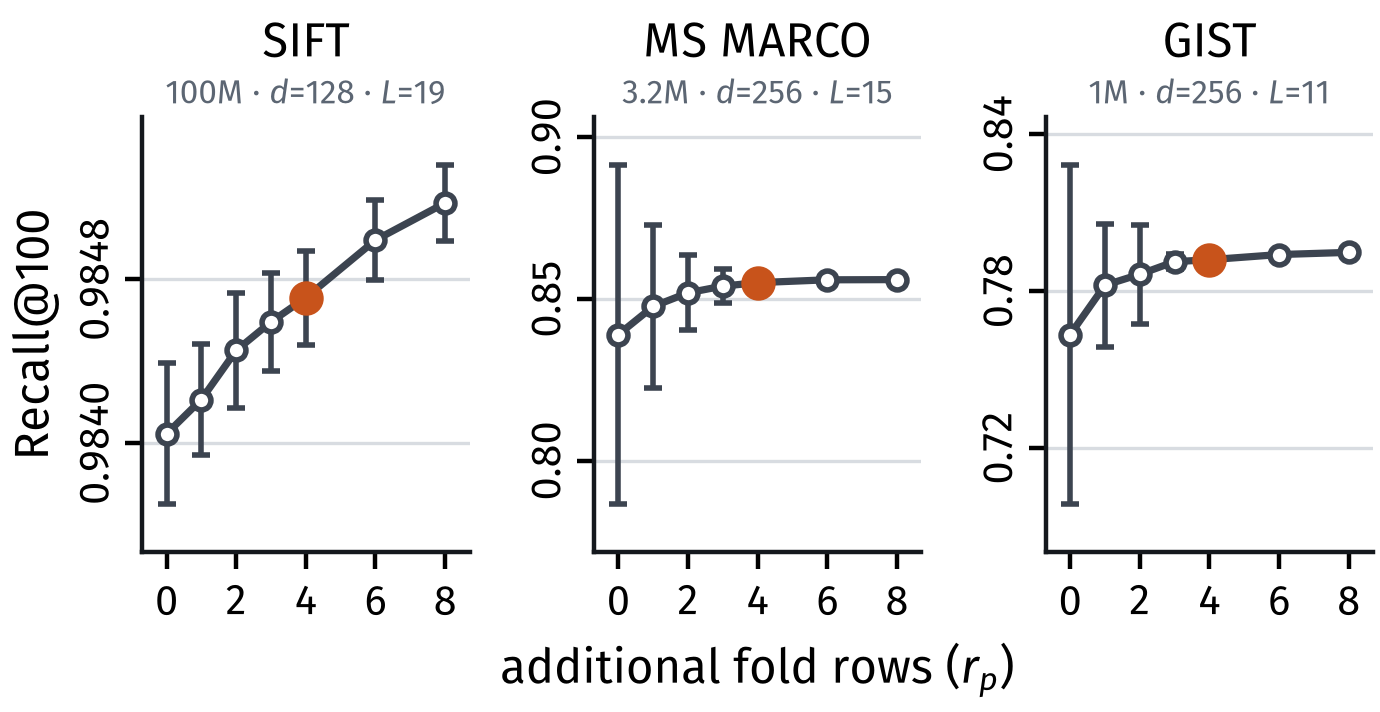}
\caption{Effect of additional random fold rows on true recall@$100$. Each point reports the mean over $20$ independently generated folding matrices, with each matrix evaluated on $1000$ queries. Error bars show one standard deviation across matrices, and the orange markers denote the deployed setting $r_p=4$.}
\label{fig:responsesize-vs-recall}
\end{figure}

Figure~\ref{fig:responsesize-vs-recall} shows that the additional rows primarily improve robustness across random folding matrices. Without additional rows, an unfavorable fold matrix can substantially degrade retrieval quality: on GIST, the worst construction achieves $0.481$ recall against a mean of $0.763$. At the deployed setting $r_p=4$, the standard deviation decreases by as much as $25.6\times$ relative to $r_p=0$, while mean recall increases by at most $0.029$. Increasing $r_p$ to $8$ reduces the standard deviation by as much as $217\times$, but improves mean recall by at most $0.0028$ beyond $r_p=4$. The effect is weakest on SIFT, where the longer $L=19$ code already exhibits little variability at $r_p=0$ and four additional rows improve mean recall by only $0.0007$.

Therefore, the response contains $R=1+L+r_p$ fold rows. \textsc{Carousel} packs two fold rows into each CKKS ciphertext by placing one in the real component and the other in the imaginary component. Each response ciphertext occupies $1.05$~MB, giving an amortized cost of $0.525$~MB per fold row. As a result, the download grows in $1.05$~MB steps, and the deployed choice $r_p=4$ adds $2.1$~MB to the response.

\subsection{Updates and Freshness}
\label{sec:eval}

Section~\ref{sec:design:updates} describes \textsc{Carousel}'s update mechanism: insertions append vectors to the database, while deletions zero the corresponding rows. Ensuring next-query freshness presents different challenges in the cached and recompute modes. We evaluate the update cost and behavior under sustained write loads in each mode below.

\smallskip\noindent\emph{Recompute mode.}
When the encoded diagonals exceed device memory capacity, \textsc{Carousel} constructs them on demand for each query. Although recomputation increases query latency, it simplifies updates because each update is an $O(d)$ write to the flat vector array. Because the next query constructs its diagonals directly from this array, it automatically observes the updated data without cache invalidation or derived-state maintenance. We measure this write path against the SIFT-100M database, which contains $100M$ 128-dimensional vectors and occupies $25.6$~GB. A single insertion takes $4.00\mu$s, corresponding to $250{,}000$ insertions per second and a deletion takes $3.25\mu$s, or $308{,}000$ deletions per second. Batching further amortizes the write overhead: batches of $10{,}000$ insertions and deletions complete in $8.8\mu$s and $7.2\mu$s, corresponding to throughputs of $1.14$ and $1.38$ billion updates per second, respectively. Concurrent updates have little effect on query performance in the recompute mode. When the writer is running on a separate CUDA stream, SIFT-100M query latency increases from $3.7244$s (without updates) to $3.7341$s at $10000$ updates per second and $3.7538$~s at $50000$ updates per second. These increases correspond to overheads of only $0.26\%$ and $0.79\%$, respectively. 

\smallskip\noindent\emph{Cached mode.}
When the encoded diagonals fit in device memory, \textsc{Carousel} caches and reuses them across queries. An update can therefore make the cached representation of the affected block stale. An update can therefore make the cached representation of the affected block stale. Append-only insertions only affect the final block, whereas random deletions can make blocks throughout the dataset stale. To preserve next-query freshness, \textsc{Carousel} marks that block as dirty and re-encodes it before it is next read. Because each vector contributes to every diagonal of its block, modifying a single row requires re-encoding the entire block. This takes $3.35$~ms on MS MARCO-256 and $3.53$~ms on GIST-256. Multiple updates to the same block before it is next read coalesce into a single re-encoding.

\begin{figure}[h]
\centering
\includegraphics[width=\columnwidth]{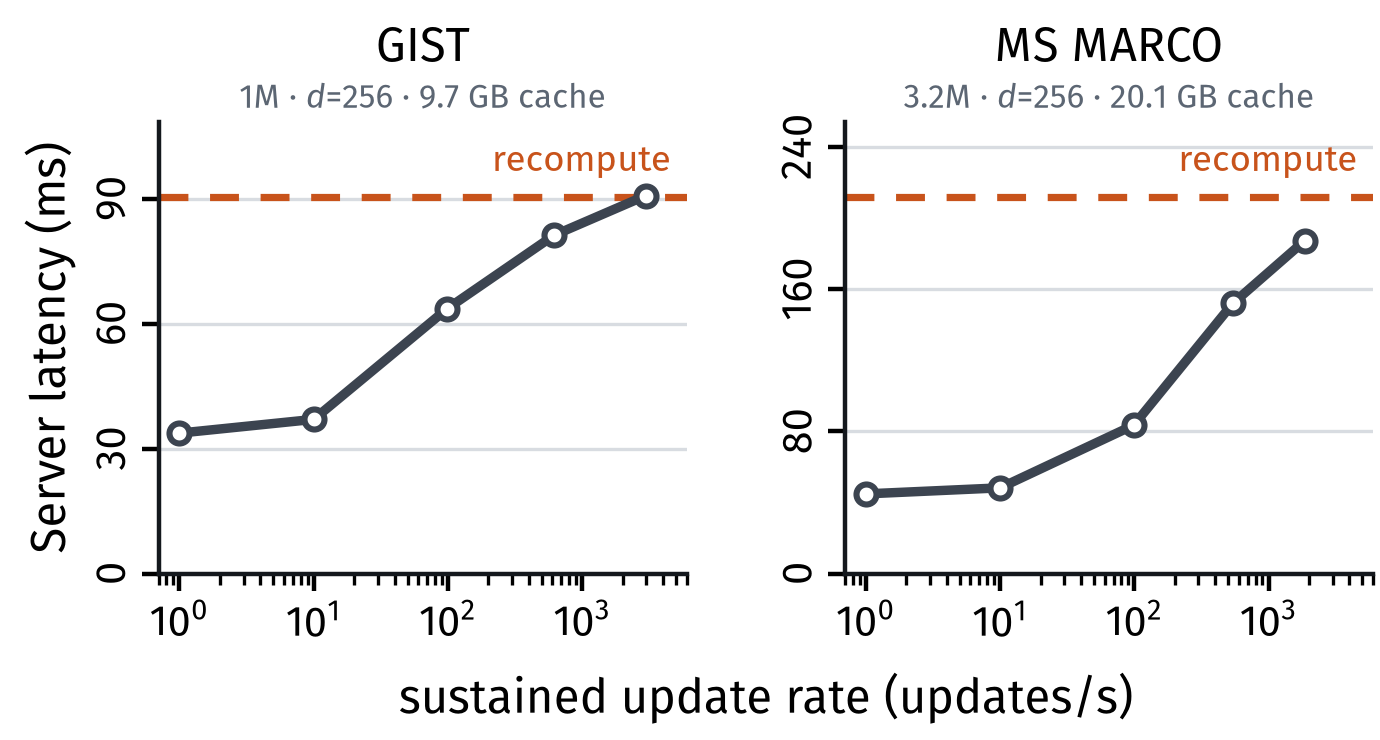}
\caption{Server latency under sustained update load in the cached mode. Each point reports the mean over $100$ queries while updates execute concurrently on a separate CUDA stream. The dashed lines show the corresponding recompute baselines.}
\label{fig:update-load}
\end{figure}

Figure~\ref{fig:update-load} shows how refreshing dirty blocks in cached mode affects query performance. Without concurrent updates, caching reduces server latency from $90.3$ to $33.8$~ms on GIST and from $211.7$ to $44.8$~ms on MS MARCO. At $10$ updates per second, cached execution remains $2.44\times$ faster on GIST and $4.39\times$ faster on MS MARCO. At $100$ updates per second, the cached mode remains $1.42\times$ and $2.54\times$ faster than the recompute mode on GIST and MS MARCO, respectively. As the update rate increases, more blocks become dirty between queries and must be re-encoded. GIST reaches the recompute baseline near $3{,}000$ updates per second. On MS MARCO, the writer saturates at $1{,}851$ updates per second, where cached execution remains $1.13\times$ faster than recomputation. Separate batch-size measurements indicate that its crossover would occur only near $3\times10^5$ updates per second. Thus, lazy re-encoding preserves the caching benefit over a wide range of update rates while maintaining next-query freshness.

For write-heavy deployments, \textsc{Carousel} can move re-encoding off the query path. Updates accumulate into batches, while a background CPU thread performs the encoding. The query thread then transfers the re-computed diagonal encodings to the GPU and swaps them into the cache. This makes query latency largely independent of the update rate. Even at $10{,}000$ updates per second, query latency remains nearly unchanged, increasing from $33.8$ to $35.1$~ms on GIST and from $44.8$ to $50.3$~ms on MS MARCO. This mode trades freshness for throughput. Queries continue using the previous cached version of an affected block until its replacement is ready, producing staleness windows of $1.9$~s on GIST and $3.7$~s on MS MARCO.

\subsection{Multi-GPU Scaling to One Billion Vectors}
\label{sec:sift1b}

\paragraph{Setting.}
We evaluate \textsc{Carousel} on the full BIGANN SIFT-1B corpus using ten GPUs. The corpus contains $10^9$ 128-dimensional vectors, which \textsc{Carousel} scans in their native representation without projection. We partition these corpus evenly across the ten GPUs, assigning $100M$ vectors to each GPU. Each GPU stores $25.6$~GB of vectors, with a total corpus size of $256$~GB across all 10 GPUs.

Each GPU receives the same encrypted query and computes a partial fold over its shard. The ten partial folds are then added homomorphically before a single response is returned. Because folding is linear, the aggregated response is equivalent to folding the complete corpus, and the client decodes once over all blocks. The sharpening exponent used is $p=128$. The diagonal encodings are recomputed for every query. The latency and recall results are measured over $1{,}000$ queries on the ten-GPU deployment.

\paragraph{Results.}

\begin{table}[h]
\centering\small
\caption{SIFT-1B on ten GPUs compared with SIFT-100M on one GPU.}
\label{tab:sift1b}
\resizebox{\columnwidth}{!}{%
\begin{tabular}{lrrr}
\toprule
& SIFT-1B & SIFT-100M & change \\
\midrule
$N$                              & $10^9$ & $10^8$ & $10\times$ \\
GPUs                             & $10$ & $1$ & $10\times$ \\
fold rows $R$                    & $29$ & $24$ & $+5$ \\
\midrule
\textbf{true recall@$100$}       & $\mathbf{0.9715}$ & $0.9845$ & $-0.0130$ \\
\textbf{true recall@$10$}        & $\mathbf{0.9699}$ & $0.9785$ & $-0.0086$ \\
recall p10 / median / p95        & $0.970$ / $0.990$ / $1.000$ & $0.970$ / $0.990$ / $1.000$ & --- \\
\midrule
\textbf{server latency}          & $\mathbf{3.765}$~s & $3.735$~s & $1.01\times$ \\
\quad per shard                  & $3.735$~s & $3.735$~s & $1.00\times$ \\
\quad query expansion            & $0.008$~s & $0.008$~s & $1.00\times$ \\
\quad encode                     & $1.993$~s & $1.993$~s & $1.00\times$ \\
\quad scan                       & $0.553$~s & $0.553$~s & $1.00\times$ \\
\quad sharpening ($7$ squarings) & $1.083$~s & $1.083$~s & $1.00\times$ \\
\quad fold                       & $0.097$~s & $0.097$~s & $1.00\times$ \\
\quad cross-shard reduction      & $0.030$~s & --- & --- \\
\midrule
client query build               & $23.3$~ms & $23.3$~ms & $1.00\times$ \\
client decrypt and decode        & $15.8$~ms & $12.0$~ms & $1.32\times$ \\
client matched filter ($C=2000$) & $375$~ms & $31$~ms & $12.1\times$ \\
\midrule
upload                           & $5.24$~MB ($1$ ct) & $5.24$~MB & $1.00\times$ \\
\textbf{download}                & $\mathbf{15.73}$~MB ($15$ ct) & $12.58$~MB & $1.25\times$ \\
\bottomrule
\end{tabular}
}
\end{table}

Table~\ref{tab:sift1b} compares SIFT-1B with the single-GPU SIFT-100M configuration. Server latency remains nearly constant after sharding. The ten-fold larger corpus takes $3.765$~s per query, only $1.01\times$ the SIFT-100M latency, because each GPU processes approximately the same number of vectors as the single-GPU SIFT-100M configuration. Processing each shard takes $3.735$~s, while aggregating the partial folds adds only $0.030$~s. Thus, on the server side, scaling from $100$ million to one billion vectors requires additional hardware rather than additional query latency.

Communication grows much more slowly than the corpus. The upload remains $5.24$~MB because the client sends one ciphertext regardless of $N$, and the server generates the remaining rotations. The download increases by only $25\%$, from $12.58$ to $15.73$~MB, because the response contains $\lceil R/2\rceil$ ciphertexts, where $R$ is the number of fold rows, and $R$ grows logarithmically with the number of blocks. Consequently, a ten-fold increase in corpus size leaves upload unchanged and increases download by only $1.25\times$. The client computation increases with corpus size. Query construction remains unchanged, while decryption and decoding increase only slightly, from $12.0$ to $15.8$~ms. The matched-filter stage grows more substantially, from $31$ to $375$~ms, because it scores candidate slots against codewords from ten times as many blocks. This measurement uses a single CPU core, and the work can be parallelized across additional cores to reduce the latency. On the other hand, reducing the number of candidate slots from $C=2000$ to $C=200$ substantially lowers the matching cost with negligible loss in recall@$10$. These two options mitigate the dominant client-side computation cost at one billion vectors.

Retrieval quality remains stable even at this scale. True recall@$100$ decreases from $0.9845$ to $0.9715$, while recall@$10$ decreases from $0.9785$ to $0.9699$. The loss is concentrated in a small subset of queries as median recall remains $0.990$, the 95th percentile remains $1.000$, and the fraction of queries achieving at least $0.95$ recall decreases from $97.3\%$ to $95.7\%$.

\subsection{Comparison with Prior Systems}
\label{sec:eval:comparison}

Table~\ref{tab:prior} compares \textsc{Carousel} with Tiptoe~\cite{tiptoe},
Wally~\cite{wally}, and Pacmann~\cite{pacmann} on the public MS MARCO corpus, with
NGT~\cite{ngt} included as a non-private quality reference. Tiptoe, Pacmann and \textsc{Carousel} hide the client's query cryptographically, whereas Wally reveals the probed cluster and bounds the leakage with differential privacy, relying on an anonymization service and a large concurrent client population. We report the performance metrics exactly as published by each system on its respective hardware. The only exception is Tiptoe because its original paper omits latency for the MS MARCO dataset, we use the measurement reported by WALLY. Since these evaluations were conducted across highly different hardware architectures, we report the metrics in their native units and refrain from drawing direct performance comparisons. \textsc{Carousel} reaches an MRR@$100$ of $0.303$ at $d=256$, the highest among the private systems and within $0.007$ of NGT. Even at a smaller projected dimension of $d=128$, \textsc{Carousel} reaches $MRR@100$ of $0.276$, higher than all prior systems. Its communication ($13.6$MB) is below Tiptoe and Pacmann, but exceeds Wally.

Tiptoe shares our setting and guarantees: a single semi-honest server, a public corpus and a single round protocol. The main difference is that while Tiptoe scores all vectors, the client only retrieves one cluster. This affects the quality of retrieval compared to \textsc{Carousel}: yielding $0.13$ against $0.303$. Tiptoe's clustering also requires the client holding $0.61$MB of centroid metadata that a corpus change invalidates, whereas \textsc{Carousel}'s client holds none. WALLY relies on this same clustering idea and suffers from the same retrieval quality problem. To escape it, WALLY allows the client to make multiple queries to retrieve different clusters. This recovers some quality but still remains low compared to \textsc{Carousel} ($0.18$ at $5$ queries against $0.303$). WALLY's only advantage is its lower communication cost, but this is due to its differential privacy approach, which requires a massive anonymity set of $5 \times 10^5$ clients to be efficient. Pacmann offers the same cryptographic guarantees but executes the search on the client, hiding each access with PIR. Therefore, the computational burden falls entirely on the client. Furthermore, Pacmann requires a preprocessing phase before any query can be issued, where the client streams a $2.7$~GB index and retains $614$~MB of hint state. In addition, each query requires $20$ sequential network rounds and $63.2$MB of communication. \textsc{Carousel} outperforms Pacmann in retrieval quality ($0.303$ against $0.266$) and matches it even when operating at a reduced dimensionality of $d=128$ (MRR@100: $0.276$). Pacmann also evaluates on SIFT, scaling to $10^{8}$ vectors with parameters tuned to reach recall@$10$ of $0.90$, at a cost of a $60$~GB preprocessing download and $3$~GB of client hint state. \textsc{Carousel} reaches recall@$10$ of $0.9785$ with $5.2$~MB of upload and $12.6$~MB of download per query.

\begin{table}[t]
\centering\small
\caption{Comparison on MS MARCO ($\approx 3.2$M documents), as reported by each system.
Tiptoe's cost figures are Wally's simulation. Wally's ranges cover probing 1 to 5 clusters.}
\label{tab:prior}
\resizebox{\columnwidth}{!}{%
\begin{tabular}{lccccc}
\toprule
& Tiptoe & Wally & Pacmann & \textbf{\textsc{Carousel}} & NGT \\
\midrule
Privacy            & crypto & DP + anon & crypto & \textbf{crypto} & none \\
Assumptions        & none & \shortstack{anon hop,\\ $5{\times}10^{5}$ clients}
                   & none & \textbf{none} & --- \\
Rounds             & $1$ & $1$ & $20$ & $\mathbf{1}$ & --- \\
\midrule
Offline pass       & none & none & $2.7$~GB & \textbf{none} & --- \\
Client state       & $0.61$~MB & $0.04$~MB & $614$~MB & \textbf{none} & --- \\
\midrule
Comm.\ (MB)        & $17.4$ & $0.56$--$2.60$ & $63.2$ & $\mathbf{13.6}$ & --- \\
Compute            & $11.0$~core-s & $0.39$--$1.50$~core-s & $0.29$~core-s
                   & $\mathbf{0.045}$~gpu-s & --- \\
\midrule
MRR@$100$          & $0.13$ & $0.12$--$0.18$ & $0.266$ & $\mathbf{0.303}$ & $0.31$ \\
\bottomrule
\end{tabular}}
\end{table}
No prior system evaluates at the billion scale. Tiptoe evaluates $360$ million embeddings on a $45$-server cluster, requiring $2.7$ seconds, $56.9$~MB of bandwidth, and $68$~MB of client-cached metadata. Pacmann scales up to $10^{8}$ vectors, requiring a $60$~GB preprocessing download and $3$~GB of hint state. Projecting Pacmann to $10^{9}$ vectors based on its scaling laws ($\tilde{O}(\sqrt{n})$ for hints, linear for preprocessing) dictates roughly $10$\,GB of state and a massive $600$\,GB download. In contrast, \textsc{Carousel} queries $10^{9}$ vectors in $3.77$ seconds on ten GPUs, downloading only $15.7$~MB with zero client state. None of the prior systems evaluate updates to the corpus. \textsc{Carousel} is the only system in this comparison that provides full cryptographic guarantees, requires only a single network round, and maintains zero corpus-derived client state. Moreover, it achieves higher retrieval accuracy than every private baseline while remaining practical at the billion scale.
\section{Discussion and Limitations}
\label{sec:discussion}

\textsc{Carousel} has four main limitations. Two of them are related to sharpening and two to deployment. Although ordinary insertions and deletions require no client action, changes to the encryption parameters require clients to re-upload their evaluation keys. The server stores three such keys per client: a multiplication key for sharpening, a rotation key for query expansion, and a conjugation key for the packing split.  When there is change in the sharpening exponent $p$, it alters the modulus-chain depth and invalidates the existing keys. As the corpus grows, folding combines scores from more blocks, increasing the accumulated bulk noise that sharpening must suppress and potentially requiring a larger $p$. Section~\ref{sec:eval:scaling} shows a transition from $p=64$ to $p=128$ beyond $29$ million vectors, after which $p=128$ remains sufficient through $1$ billion vectors. This is \textsc{Carousel}'s only nonincremental reconfiguration. Across our SIFT scaling experiment, it occurs only once, requires no re-indexing, and transfers less than $32$~MB of replacement keys per client. A second limitation is that sharpening orders scores by absolute magnitude rather than signed value. Because the exponent is an even power of two, a strongly negative score could become heavy and returned as a match. In practice, this issue does not arise, as the most negative score a query encounters is consistently much smaller in magnitude than its largest positive match. While using an odd exponent would preserve the sign, it requires one additional ciphertext multiplication and increases the overall multiplicative depth. This configuration is omitted as no corpus requires it.

The remaining constraints are deployment-driven. Device memory limits whether encoded diagonals can be cached, forcing larger configurations to use recompute mode. Client decoding scales with the block count $B$ and the candidate budget $C$, the number of slot candidates the client evaluates. Matched-filter latency increases from $31$~ms at $10^{8}$ vectors to $375$~ms at $10^{9}$, where it dominates client cost. Reducing $C$ lowers this latency but can reduce recall.

\section{Conclusion}
\textsc{Carousel} shows that single-round private nearest-neighbor search with FHE is practical at billion-vector scale. Prior systems avoid fully homomorphic evaluation and instead use linearly or somewhat homomorphic encryption.By using GPU-accelerated CKKS, \textsc{Carousel} makes exhaustive FHE scanning practical and acheives high recall by not discarding potential candidates. However ranking those scores still remains expensive, so \textsc{Carousel} introduces sharpening and folding to complete the private search pipeline efficiently.  \textsc{Carousel} also keeps no corpus-derived state on the client. The client holds only its keys and a public projection, so a change to the corpus invalidates nothing it holds. This makes updates practical, whereas prior work in this setting either treats them as an open problem or briefly discusses them as an expensive operation. Overall, this design achieves the highest retrieval quality among private search systems, matching that of a non-private index, and is the only one scalable upto a billion vectors. \textsc{Carousel} could be applied to a wide range of applications including private web search, recommendation, and retrieval-augmented generation.

\begin{acks}
This work was supported by the Air Force Research Laboratory /
Information and Spectrum Directorate (AFRL/RF), Rome, NY.
Any statements or claims made in this publication are solely
those of the authors and do not represent the views or policies
of the United States Government.
\end{acks}

\clearpage

\balance

\bibliographystyle{ACM-Reference-Format}
\bibliography{reference}


\end{document}